\documentclass[final,1p,times]{elsarticle}

\usepackage{amssymb}

\usepackage{lineno}
\journal{Journal of Computational Physics}
\usepackage[utf8]{inputenc}
\usepackage[T1]{fontenc}
\usepackage{subfiles}
\usepackage{amsmath,amssymb,bm}
\usepackage{amsthm}
\usepackage{hyperref}
\usepackage{times}
\usepackage{framed}
\usepackage{graphicx}
\usepackage{svg}
\usepackage{verbatim}	
\usepackage{breqn} 
\usepackage{listings}
\usepackage{bbm}
\usepackage{subcaption}
\usepackage{booktabs}

\usepackage{xcolor}
\usepackage[ruled,vlined,linesnumbered ]{algorithm2e} 

\usepackage{enumitem}
\usepackage{pifont}

\usepackage{cleveref} 
\usepackage{titlesec}
\usepackage{setspace}

\newcommand{\bc}{\begin{center}}
\newcommand{\ec}{\end{center}}

\newcommand{\bmat}{\begin{pmatrix}}
\newcommand{\emat}{\end{pmatrix}}
\newcommand{\bsmat}{\left(\begin{smallmatrix}}
\newcommand{\esmat}{\end{smallmatrix}\right)}
\newcommand{\bes}{\begin{equation}\begin{split}}
\newcommand{\ees}{\end{split}\end{equation}}

\newcommand{\refeq}[1]{Equation (\ref{#1})} 
\newcommand{\refsec}[1]{Section (\ref{#1})}
\newcommand{\reffig}[1]{Fig. (\ref{#1})}

\newcommand\be{\begin{equation}}
\newcommand\ee{\end{equation}}

\newcommand\bea{\begin{eqnarray*}}
\newcommand\eea{\end{eqnarray*}}

\newcommand{\bd}{\begin{description}}
\newcommand{\ed}{\end{description}}
\newcommand{\bi}{\begin{itemize}}
\newcommand{\ei}{\end{itemize}}
\newcommand{\bb}{\begin{block}}
	\newcommand{\eb}{\end{block}}
\newcommand{\pa}{\partial}

\newcommand{\bs}{\boldsymbol}

\newcommand\calF{\mathcal{F}}
\newcommand\calG{\mathcal{G}}

\newcommand\calL{\mathcal{L}}

\newcommand\calI{\mathcal{I}}
\newcommand\calD{\mathcal{D}}
\newcommand\calN{\mathcal{N}}

\newcommand\RR{\mathbb{R}}

\newcommand{\bxi}{\boldsymbol{\xi}}

\DeclareMathAlphabet\mathbfcal{OMS}{cmsy}{b}{n}

\def\bt{\bs{\theta}}

\def\BState{\State\hskip-\ALG@thistlm}

\newcommand{\stkout}[1]{\ifmmode\text{\sout{\ensuremath{#1}}}\else\sout{#1}\fi}

\newcommand{\update}[1]{\textcolor{black}{#1}}

\newcommand{\epsTau}{\bm \epsilon_{\tau}}

\newcommand{\tauTheta}{\bm{\tau}_{\bm \theta}}
\newcommand{\taub}{\bm{\tau}}

\newcommand{\zobs}{\hat{\bm{z}}}
\newcommand{\z}{\bm z}
\newcommand{\thetaMAP}{\bm{\theta}_{MAP}}
\newcommand{\LambdaMAP}{\bm{\Lambda}_{MAP}}
\newcommand{\ETau}{\bm E_{\tau}}

\begin{document}

\begin{frontmatter}



\title{A probabilistic, data-driven closure model for RANS
simulations with aleatoric,  model uncertainty}





\author[label1]{Atul Agrawal}
\ead{atul.agrawal@tum.de}
\affiliation[label1]{organization={Technical University of Munich},
            addressline={Professorship of Data-driven Materials Modeling, School of
Engineering and Design, Boltzmannstr. 15}, 
            city={85748 Garching},
            country={Germany}}
            
\author[label1,label2]{Phaedon-Stelios Koutsourelakis\corref{cor1}}

\cortext[cor1]{Corresponding author}
\ead{p.s.koutsourelakis@tum.de}

\affiliation[label2]{organization={Munich Data Science Institute (MDSI - Core member)},
            city={Garching},
            country={Germany}}

\begin{abstract}
We propose a data-driven, closure model for Reynolds-averaged Navier-Stokes (RANS) simulations that incorporates aleatoric, model uncertainty. 
The proposed closure consists of two parts. A parametric one, which  utilizes  previously proposed, neural-network-based  tensor basis functions dependent on the rate of strain and rotation tensor invariants. This is complemented by latent, random  variables which account for aleatoric model errors. 
A fully Bayesian formulation is proposed, combined with  a sparsity-inducing prior in order to identify regions in the problem domain  where the parametric closure is insufficient and where stochastic corrections to the Reynolds stress tensor are needed. Training is performed using sparse, indirect data, such as mean velocities and pressures, in contrast to  the majority of alternatives  that require direct Reynolds stress data. For inference and learning, a Stochastic Variational Inference scheme is employed, which is based on   Monte Carlo estimates of the pertinent objective in conjunction with the reparametrization trick. This necessitates derivatives of the output of the RANS solver, for which we developed an adjoint-based formulation. In this manner, the parametric sensitivities from the differentiable solver can be combined with the built-in, automatic differentiation capability of the neural network library in order to  enable an end-to-end differentiable framework. We demonstrate the capability of the proposed model  to produce accurate, probabilistic, predictive estimates  for all flow quantities, even in regions where model errors are present, on a separated flow in the backward-facing step benchmark problem. 
\end{abstract}


\end{frontmatter}

\section{Introduction}
\label{sec:intro}

Turbulence is ubiquitous in fluid flows 
and of importance to  a vast range of applications such as  aircraft design, climate and ocean modeling. It has challenged and intrigued scientists and artists  for  centuries \cite{doi:10.1146/annurev-fluid-022620-122816}.
In the context of the Navier-Stokes equations, the most accurate numerical solution strategy for turbulent flows is offered by Direct Numerical Simulation (DNS), which aims at fully resolving all scales of motion. While this simulation method yields impeccable results, it is prohibitively expensive in terms of  computational cost due to the very fine discretizations needed  which scale as $\mathcal{O}(\mathrm{Re}^{11/4})$ \cite{pope2000turbulent}.
  Reynolds-averaged Navier-Stokes (RANS) models offer a much more efficient alternative for predicting mean flow quantities. They  represent the  industry standard which is expected to remain the case  in the coming decades \cite{Slotnick2014}. Their predictive accuracy however hinges upon the closure model adopted.

 The greater availability of computational resources and the development of scalable learning frameworks in the field  of machine learning have had a significant  impact in  computational fluid mechanics as well \cite{brunton2020machine,vinuesa_emerging_2022,Lucor2022}. 
  Data-driven closures for RANS have revitalized turbulence modeling \cite{duraisamy_perspectives_2021}, and a comprehensive reviews can be found in \cite{duraisamy2019turbulence,brunton2020machine}. The construction of such  closure models consists of two steps: (i) postulating a model form ansatz; and (ii)  fitting/learning/inferring model parameters on the basis of the available data. Pertinent  approaches  have focused on learning model coefficients of a given turbulence model \cite{Oliver2011} (often with statistical inference), on modeling of correction or source terms for an existing turbulence model \cite{parish2016paradigm,singh2017machine,tracey2015machine,xiao2016quantifying,zhang2015machine} and on directly modeling the Reynolds stress (RS) tensor \cite{ling2016reynolds,Ling2016,kaandorp2020data,wang2017comprehensive}  with symbolic regression \cite{schmelzer2020discovery} or neural networks \cite{ling2016reynolds,kaandorp2020data,zhang_ensemble_2022} or Gaussian Processes \cite{zhang2015machine} or Random Forests \cite{Ling2016,wang2017comprehensive}.
Of particular relevance to the present study is the work of \cite{ling2016reynolds} wherein they use the non-linear eddy viscosity model (NLEVM) \cite{pope1975more} to capture  the anisotropic part of the RS tensor using an integrity tensor basis and a deep neural network employing  local, invariant  flow features.
 This model owing to its guaranteed Galilean invariance  found a wider utilization \cite{kaandorp2020data,geneva2019quantifying,zhang_ensemble_2022}. 

%
%
%
In most of the methods discussed above, data-based training is performed in a \emph{non-intrusive} manner, i.e., without involving the RANS solver in the training process. 
The major shortcomings of such a strategy (which we attempt to address in the present paper) are two-fold. Firstly inconsistency  issues, which can arise between the data-driven model and the baseline turbulence model (e.g., $k-\epsilon$) \cite{taghizadeh2020turbulence,duraisamy_perspectives_2021}. 
\cite{thompson2016methodology} showed that even substituting RS fields from reputable DNS databases may not lead to satisfactory prediction of the velocity field,  and \cite{wu2019reynolds} investigated  the ill-conditioning that arises in the RANS equations, when employing  data-driven models that treat the Reynolds stress as an explicit source term. 
This ill-conditioning can be amplified within each iteration, thus potentially leading to divergence during the solution procedure.
Secondly, such models rely on  full-field Reynolds stress training data, which are only  available   when  high-fidelity simulations such as DNS/Large-Eddy Simulations (LES) are used. Unfortunately, such high-fidelity simulations due to their expense are  limited to simple geometries and low Reynolds numbers.

In order to address these limitations, we advocate  incorporating the   RANS model in the training process. This enables one to use indirect data  (e.g., mean velocities and pressure) obtained from higher-fidelity simulations or experiments as well as direct data (i.e. RS tensor observables) if this is available.  In the subsequent discussions, we will refer to such a training strategy as "model-consistent learning" \cite{duraisamy_perspectives_2021}. It necessitates the  solution of a high-dimensional inverse problem that minimizes a discrepancy measure between the RANS solver's output (mean velocities and pressure) and the observables (e.g. mean fields from LES/DNS). 
\update{As pointed out in \cite{duraisamy_perspectives_2021}, model-consistent training or simulation-based Inference \cite{cranmer2020frontier} benefits from the differentiability of the solver, as it provides  derivatives of the outputs with respect to the tunable parameters that can significantly expedite the learning/inference process.}


In recent years there has been a concerted effort towards  developing differentiable CFD solvers \cite{bezgin2021fully,list2022learned,um_solver---loop_2021, dmitrii_kochkov_machine_2021} in Auto-Differentiation (AD) enabled modules like PyTorch, Tensorflow, JAX, Julia. 
To the best of the authors' knowledge, this has not been accomplished yet for RANS solvers.  One way to enable the computation of parametric sensitivities   is by developing  adjoint solvers \cite{giles2003algorithm,giles2000introduction}, which  are  commonly used in the context of aerodynamic shape optimization \cite{jameson1988aerodynamic}. 
Such adjoint-based modules have also been employed to infer a spatial, corrective field for transport equations \cite{parish2015quantification,parish2016paradigm,singh2017machine} and  Reynolds stresses \cite{xiao2016quantifying}.
Recently, \cite{holland_towards_nodate, bidar_open-source_2022} tried to learn a corrective,  multiplicative field in the production term of the Spalart–Allmaras transport model. This is based on an alternative approach outlined in  \cite{parish2016paradigm}, in which empirical correction terms for the turbulence transport equations are learned while retaining a traditional linear eddy viscosity model (LEVM) for the closure. 
\cite{brenner2022efficient} used adjoints  to recover a spatially varying eddy viscosity correction factor from sparsely distributed training data, but they also retained the LEVM assumption. 
\update{More recently, researchers performed model-consistent or CFD-driven training by involving the solver in the training process. \cite{saidi2022cfd,zhao2020rans} used gradient-free algorithms to perform symbolic identification of the explicit algebraic Reynolds stress models (ARSM)}
%
%
, \cite{strofer2021end} (with adjoint methods) and \cite{zhang_ensemble_2022} (with ensemble methods) combined the RANS solver and a NLEVM-based neural network  proposed by \cite{ling2016reynolds} in order to learn the  model closure. However, they did  not account for  potential model errors  in the closure equation which may arise due to the reasons discussed in the next paragraph. 





%


We argue that even in model-consistent training,  a discrepancy in the learnt RS closure model can arise due to the fact that a) the parametric, functional form employed may be insufficient to represent the underlying model\footnote{For example the models based on the Boussinesq hypothesis will fail to capture the flow features driven by the anisotropy of the Reynolds stresses and this intrinsic deficiency cannot be remedied by the calibration of the model coefficients  with data.},
and b) the flow features which are used as  input in the closure relation and which are generally restricted to each point in the problem domain (locality/Markovianity assumption \cite{parish2017non}), might not contain enough information to predict the optimal RS tensor leading to  irrecoverable loss of information. Hence irrespective of the type and amount of training data available, there could be  {\em aleatoric} uncertainty in the closure model that needs to be quantified and propagated in the predictive estimates. 
%
We note that much fewer  efforts have been directed towards quantifying  uncertainties in  RANS turbulence models.
Earlier, parametric approaches  broadly explored the uncertainties in the model choices \update{\cite{Oliver2011}} ( i.e., uncertainty involved in choosing the best model among a class of competing models, e.g., $k-\omega, k-\epsilon$) and their respective model coefficients \update{\cite{schaefer2017uncertainty}}. 
Recently the shortcomings of the parametric closure models have been recognized by the turbulence modeling community \update{\cite{xiao2019quantification, wang2016quantification}}.
In light of this, various non-parametric approaches  have targeted model-form uncertainty whereby uncertainties are  directly introduced into the turbulent transport equations or the modeled terms such as the Reynolds stress \cite{geneva2019quantifying} or eddy viscosity. Such formulations allow for more general estimates of the model inadequacy than the parametric approaches.
Researchers have also tried perturbing the eigenvalues \cite{emory2013modeling,gorle2013framework,edeling2018data}, transport  eigenvectors \cite{thompson2019eigenvector} or the tensor invariants.
\cite{wu2017priori} used kernel density estimates to predict the confidence of  data-driven models, but it is limited to the prediction of the anisotropic stress and fails to provide any  probabilistic bounds. \cite{geneva2019quantifying} tries to address this issue by incorporating a Bayesian formulation in order  to quantify epistemic  
uncertainty and then propagating it to quantities of interest  like pressure and velocity. 
%
%
For a comprehensive review of modeling uncertainties in the RANS models  the reader is directed to \cite{xiao2019quantification}.

In order to  address the aforementioned  limitations, we propose a novel probabilistic, model-consistent, data-driven differential framework. The framework enables learning of a NLEVM-based, RS model in a model-consistent way using a differentiable RANS solver, with mean field observables (velocities and/or pressure). To the authors' knowledge, uncertainty quantification has not been addressed for data-driven turbulence model training with indirect observations. We propose to augment the parametric closure model of the RS tensor by a stochastic discrepancy tensor to quantify model errors at different parts of the problem domain. With the introduction of the stochastic discrepancy tensor, we advocate a probabilistic formulation  for  the associated  inverse problem, which provides a  superior setting as it is capable of quantifying predictive uncertainties which are unavoidable when any sort of model/dimensionality reduction is pursued and when the  model (or its closure) is learned from finite data  \cite{koutsourelakis2016big}. 
%
%
To achieve the desired goals, the proposed framework employs the following major elements:

\begin{itemize}
	\item A discrete, adjoint-based differentiable RANS solver to enable model-consistent, gradient-based learning (\refsec{sec:methods}, \refsec{sec:learning}).
	\item The RS closure model consists of a parametric part that is expressed with an invariant neural network as proposed in \cite{ling2016reynolds}  (\refsec{sec:RS model}), to which a {\em stochastic} discrepancy tensor field is added in order to  account for the insufficiency of the parametric  part (\refsec{sec:discrepancy tensor}).
	\item A fully Bayesian formulation that enables the quantification of epistemic uncertainties and their propagation to the predictive estimates (\refsec{sec:learning}, \refsec{sec:prediction})
	\item This is combined with a {\em sparsity-inducing} prior  model that activates the discrepancy term 
	only in regions of the problem domain where the parametric model is insufficient (\refsec{sec:discrepancy tensor}). 
\end{itemize}
%
%


The structure of the rest of the  paper is as follows. \refsec{sec:method_general} presents the governing equations and their discretization, the closure model proposed consisting of the parametric part and the stochastic corrections provided by latent variables introduced. We also present associated prior and posterior densities, a stochastic Variational Inference scheme that was employed for identifying model parameters and variables as well as the computation of predictive estimates with the trained model.
Finally, \refsec{sec:numerical Illustrations} discusses the implementation aspects and demonstrates the accuracy and efficacy of the proposed framework in the backward-facing step test case \cite{gresho1993steady}, where the linear eddy viscosity models are known to fail. We compare our results with LES reference values and the $k-\epsilon$ model, which is arguably the most commonly used RANS model.  In \refsec{sec:conclusion}, we summarize our findings and discuss limitations and potential enhancements.


\section{Methodology}
\label{sec:method_general}
\subsection{Problem Statement} \label{sec:problem formulation}

\subsubsection{Reynolds-Averaged Navier-Stokes (RANS) equations}

The Navier-Stokes equations for incompressible flows of  Newtonian fluids are given by \update{(in indicial notation)}:
\begin{align} \label{eq:NS}
	\frac{\pa U_i}{\pa t} + \frac{\pa}{\pa x_j}(U_iU_j) &= \nu \frac{\pa^2 U_i}{\pa x_j \pa x_j} - \frac{1}{\rho}\frac{\pa P}{\pa x_i}, \\
	\frac{\pa U_j}{\pa x_j} & = 0,	
\end{align}
%
where \update{$i,j$ are free and dummy indices respectively taking values $1,2,3$ and } $U_i$, $P$, $t$, $x_j$, $\nu$ and $\rho$
 represent  the flow velocity, pressure, time, spatial coordinates, the dynamic viscosity and the density of the fluid 
 respectively.  
The non-linearity of the convective term $\frac{\pa}{\pa x_j}(U_iU_j)$ gives rise to  chaotic solutions beyond a critical value of the 
\textit{ Reynolds number} $Re$. 
This necessitates very fine spatio-temporal discretizations in order to capture the salient scales. 
Such brute-force, fully-resolved simulations, commonly referred to as Direct Numerical Simulations (DNS), can become   prohibitively expensive, particularly as   $Re$ increases. 

The velocity field can be decomposed into its time-averaged  (or mean) part  $u$ and the part corresponding, to generally fast,  fluctuations  $\tilde{u}$ as:
\begin{align}
	U_i(\bm{x},t)&=u_i(\bm{x}) + \tilde{u}_i(\bm{x},t), \\ \text{where,} ~~~ u_i(\bm{x}) &= \left<U_i(\bm{x},t)\right>=  \lim_{T\to \infty} \frac{1}{T} \int_{0}^{T}U_i(\bm{x},t)~dt.
	\label{eq:udef}
\end{align}
Similarly the pressure field is also decomposed as
\begin{align}
	P(\bm{x},t)&=p(\bm{x}) + \tilde{p}(\bm{x},t), \\ \text{where} ~~~ p(\bm{x}) &= \left<P(\bm{x},t)\right>= \lim_{T\to \infty} \frac{1}{T} \int_{0}^{T}P(\bm{x},t)dt.
	\label{eq:pdef}
\end{align}

%
%
Substituting these decompositions into the Navier-Stokes equations (\refeq{eq:NS}) and applying time-averaging results in the Reynolds-averaged Navier-Stokes (RANS) equations \cite{pope2000turbulent,alfonsi_reynolds-averaged_2009}, i.e.:
%
\begin{align}
	u_j \frac{\pa u_i}{\pa x_j} -\nu \frac{\pa^2 u_i}{\pa x_j \pa x_j} +\frac{1}{\rho} \frac{\partial p}{\partial x_i}&= -  \frac{\partial \left<\tilde{u}_i\tilde{u}_j\right>}{\partial x_j}, \label{eq:RANS_equation}\\
	\frac{\pa u_i}{\pa x_i} & = 0,
	\label{eq:RANScons}
\end{align}
%
%
%
%
%
where $\left< \cdot \right>$ denotes the time average of the arguments as in \refeq{eq:udef} or \refeq{eq:pdef}. 
In several engineering applications involving  turbulent flows, the quantities of interest depend upon the time-averaged quantities. These  can be obtained by solving the RANS equations which in general implies a much lower computational cost than DNS. 

\subsubsection{The closure problem}

%
The RANS equations are unfortunately {\em unclosed} as they depend on the cross-correlation of the fluctuating velocity components, commonly referred to as the Reynolds-Stress (RS) tensor $\bm{\tau}_{RS}$:
%
\begin{align}
	\bm{\tau}_{RS} = - \left<\tilde{u}_i\tilde{u}_j\right>.  
\end{align}
%
The goal of pertinent efforts is therefore to devise appropriate closure models where   the RS tensor $\bm{\tau}_{RS}$ is expressed as a function as the primary state variables in the RANS equations i.e. the time-averaged flow quantities.
\update{Closure models are of three types: (i) Functional, which use physical insight to construct the closure; (ii) Structural, which use mathematical tools; and (iii) Data-driven, which employ experimental/simulation data \cite{ahmed_closures_2021}. For a comprehensive review, the reader is directed to \cite{san_neural_2017, ahmed_closures_2021,snyder_reduced_2022}.}
%
Classically, turbulence models are devised to represent higher-order moments of the velocity fluctuations in terms of lower-order moments. This can be done directly, as in the case of the eddy-viscosity models, or indirectly, as in the case of models based on the solution of additional partial differential equations \cite{pope2000turbulent}. 

The most commonly employed strategy is based on  the linear-eddy-viscosity-model (LEVM), which uses the Boussinesq approximation according to which $\bm{\tau}_{RS}$  is  expressed as:
\begin{align}\label{eq:LEVM model}
	\taub_{LEVM} = \frac{2}{3} k \bm{I} - 2 \nu_{t} \bm{\bar{S}},
\end{align}
where $\nu_{t}$ is the eddy viscosity, $\bm{\bar{S}}=\frac{1}{2}\left(\bm{\nabla} \bm{u}+ \bm{\nabla} \bm{u}^T \right)$ is the mean strain-rate tensor, $\bm{I}$ is the second order identity tensor, and $k = - \frac{1}{2} \text{tr}(\taub_{RS})$ is the turbulent kinetic energy. 
The eddy viscosity is computed after solving the equation(s) for the turbulent flow quantities such as  the turbulent kinetic energy $k$ and the turbulent energy dissipation $\epsilon$ (e.g. the $k-\epsilon$ model \cite{Launder1974}), or the specific dissipation $\omega$ (e.g. the $k-\omega$ \cite{Wilcox2008}).
%
Although the Boussinesq approximation  provides accurate results  for a range of flows, it can  give rise to predictive  inaccuracies which are particularly prominent when trying to capture flows with significant curvatures, recirculation zones, separation, reattachment, anisotropy, etc  \cite{pope2000turbulent,wilcox1998turbulence}. 
%
%
%
%
%
\update{Attempts to overcome this weakness have been made in the form of nonlinear eddy viscosity models (e.g., \cite{Speziale1987,pope1975more, craft1996development}), Reynolds-stress transport models (e.g., \cite{launder1975progress}) and ARSM (e.g.,\cite{gatski1993explicit,girimaji1996fully}). These models have not received widespread attention because they lack the  robustness of LEVM and involve more parameters that need to be calibrated.}

\subsection{Probabilistic, model-consistent data-driven differential framework}\label{sec:methods}

Upon discretisation  using e.g. a finite element scheme (\ref{app:RANS solver details}),  one can express the RANS equations   (\refeq{eq:RANS_equation}) in residual form as:
\begin{align}
	\mathcal{G}(\z)& = \bm{B} \taub\\
	\text{or, ~~} \mathcal{R}(\z; \taub):= \mathcal{G}(\z)& - \bm{B} \taub = 0, \label{eq:residual_tau_theta}
\end{align}
where $\z = [\bm u, \update{p}]^T$ summarily denotes the discretized velocity $\bm{u}$ and pressure $p$ fields and $\taub$ the discretized RS field. 
E.g. for a two-dimensional flow domain $\z \in \mathbb{R}^{N \times 3}$, $\bm{\tau} \in \RR^{N \times 3}$  where $N$ is the number of grid points.
The discretization scheme employed and other implementation details are discussed in  \ref{app:RANS solver details}.
We denote with $\mathcal{G}$  the discretized,  non-linear operator accounting for the  advective and diffusive terms on the left-hand side of \refeq{eq:RANS_equation} as well as the conservation of mass in \refeq{eq:RANScons}, and with $\bm{B}$ the matrix (i.e. linear operator) arising from the divergence term on the right-hand side of \refeq{eq:RANS_equation}.

Traditional, data-driven strategies postulate a closure e.g. $\bm{\tau}_{\bt}(\z)$ (or most often $\bm{\tau}_{\bt}(\bm{u})$) dependent on some tunable parameters $\bt$, which they determine either by assuming  that reference Reynolds-stress data is available from DNS simulations (or in general, from higher-fidelity models such as LES)  
or by employing experimental or simulation-based data of the mean velocities/pressures i.e. of $\z$. The former scenario which is referred to as {\em model regression} \cite{10.1063/5.0061577} has received significant attention in the past (e.q. \cite{geneva2019quantifying,ling2016reynolds,wang2017physics,kaandorp2020data}). Apart from the heavier data requirements, it does not guarantee that the trained model would yield accurate predictions of $\z$ \cite{thompson2016methodology} as even small errors in $\bm{\tau}$ might get amplified when solving \refeq{eq:residual_tau_theta}. 
The second setting, referred to as {\em trajectory regression} in \cite{10.1063/5.0061577}, might be able to make use of  indirect and noisy observations but is  much more cumbersome as repeated model evaluations and parametric sensitivities, i.e. a differentiable solver, are needed for training.

Critical to  any data-driven model is  its ability to generalize i.e. to produce accurate predictions under different flow scenaria. On one hand this depends on the training data available but on the other, on incorporating a priori available domain knowledge. The latter can attain various forms and certainly includes known invariances or equivariances that characterize the associated maps. 
Apart from this and the particulars of the parameterized model form, a critical aspect pertains to uncertainty quantification. We distinguish between parametric  and model uncertainty. The former is of epistemic origin and  has been extensively studied 
\update{(e.g., \cite{Oliver2011,schaefer2017uncertainty,xiao2019quantification}).}
Bayesian formulations offer a rigorous manner for quantifying it and ultimately propagating it in the predictive estimates in the form of the predictive posterior. We note however  that in the limit of infinite data, the posterior of the model parameters $\bt$ (no matter what these are or represent) would collapse to a Dirac-delta i.e. point-estimates for $\bt$ would be obtained. This false lack of uncertainty does not imply that the  model employed is perfect as the true (unknown) closure might attain a form not contained in the parametric family used or in the features of $\z$  that appear in the input (e.g. even though  all models proposed employ a locality assumption in the closure equations, non-local features of $\bm{u}$ might be needed).

The issue of model uncertainty in the closure equations which is of an aleatoric nature, has been much less studied and represents the main contribution of this work. In particular, we augment the parametric closure model $\bm{\tau}_{\bt}(\bm{u})$  with a set of latent (i.e. unobserved) random variables $\bm{\epsilon}_{\tau}$ which are embedded in the model equations and which  quantify model discrepancies at each grid point. 
In reference to the discretized RS vector $\bm{\tau}$ in \refeq{eq:residual_tau_theta}, we propose: 
\begin{align}
\bm{\tau}=\bm{\tau}_{\bt}(\bm{u})+\bm{\epsilon}_{\tau}.
\label{eq:taudecomp}
\end{align}
We emphasize the difference between model parameters $\bt$ and the random variables $\bm{\epsilon}_{\tau}$. While both are informed by the data, the latter remain random even in the limit of infinite data.
As we explain in the sequel, we advocate a fully Bayesian formulation that employs indirect observations of the velocities/pressures. These are combined with appropriate sparsity-inducing priors which can turn-off model discrepancy terms when the parametric model is deemed to provide an adequate fit. In this manner, the regions of the problem domain where the closure is most problematic are identified while probabilistic, predictive estimates  are always obtained.
In particular, in  \refsec{sec:RS model}  the parametric part of the closure model is discussed. In \refsec{sec:discrepancy tensor}  the proposed, stochastic, discrepancy tensor is presented. In \refsec{sec:model parametrization} prior and posterior densities are discussed and in  \refsec{sec:learning} the corresponding inference and learning algorithms are introduced. Finally  in \refsec{sec:prediction},  the computation of predictive estimates using the trained model is discussed.

\subsubsection{Parametric RS  model}\label{sec:RS model}

In this section we discuss the parametric part, i.e. $\bm{\tau}_{\bt}(\bm{u})$ in the closure model of \refeq{eq:taudecomp}.
As this represents a vector containing its values at various grid points over the problem domain, the ensuing discussion and equations should be interpreted as per grid point.
We note that the most popular LEVM model (\refeq{eq:LEVM model})
assumes that the anisotropic part of the $\bm{\tau}_{LEVM}$, is linearly related to the mean strain rate tensor $\bm{\bar{S}}$.
%
This linear relation assumption restricts the model to attain a small subset of all the possible states of turbulence.
This subset is referred to as the plane 
strain line \cite{Iaccarino2017}. Experimental and DNS data show that turbulent flows explore large regions of the domain of realizable turbulence states.

In the present work, we make use of  the invariant neural network architecture proposed by \cite{ling2016reynolds} which relates  the anisotropic part of the RS tensor with  the symmetric and antisymmetric  components of the velocity gradient tensor. By using  tensor invariants, the neural network is able to achieve both Galilean invariance as well as rotational invariance. The Navier-Stokes equations are Galilean-invariant, i.e. they remain unchanged for all inertial frames of reference. The theoretical foundation of this neural network lies in  the Non-Linear Eddy Viscosity Model (NLEVM) proposed by \cite{pope1975more} and has been used in several studies \cite{geneva2019quantifying,xiao2016quantifying,kaandorp2020data}. By employing  barycentric realizability maps \cite{gorle2013framework,emory2013modeling, mishra_uncertainty_2017, Iaccarino2017,thompson2019eigenvector}, it was shown in  \cite{kaandorp2020data}   that  this architecture  overcomes the plane strain line restriction and can explore  other realizable states.

In the model proposed by \cite{pope1975more}, the normalized anisotropic tensor of the R-S was given by  $\bm b:= \bm b(\bm{S}, \bm{\Omega})$,  which was a function of the normalized mean rate of strain tensor $\bm{S}$ and the rotation tensor $\bm{\Omega}$, i.e.:
\begin{align}\label{eq:NLEVM}
	\taub = 2k \bm{b} + \frac{2k}{3}\bm{I}; \quad \bm{S}=\frac{1}{2} \frac{k}{\epsilon}\left(\bm{\nabla} \bm{u}+ \bm{\nabla} \bm{u}^T \right) \quad \bm{\Omega}=\frac{1}{2}\frac{k}{\epsilon}\left(\bm{\nabla} \bm{u} - \bm{\nabla} \bm{u}^T \right).
\end{align}
Through the application of Cayley-Hamilton theorem \cite{pope1975more},  the following general expression for the anisotropy tensor $\bm b$ was adopted:
\begin{align}\label{eq:NLEVM_b}
	\bm{b} = \sum_{k=1}^{10}G^{(k)}(\underbrace{\calI_1 ... \calI_5}_{\substack{Scalar\\Invariants}}) \bm{\mathcal{T}}^{(k)},
\end{align}
where:
\begin{align}
	\calI_1 = \mathrm{tr} (\bm{S}^2), \quad
	\calI_2 = \mathrm{tr} (\bm{\Omega}^2), \quad
	\calI_3 = \mathrm{tr} (\bm{S}^3), \quad
	\calI_4 = \mathrm{tr} (\bm{\Omega}^2 \bm{S}), \quad
	\calI_5 = \mathrm{tr} (\bm{\Omega}^2 \bm{S}^2),
\end{align}
and $\bm{\mathcal{T}}^{(k)}$ are the symmetric tensor basis  functions (the complete set is listed in  Table \ref{table:tensor_basis}). The coefficients $G^{(i)}$ are  scalar, non-linear functions which depend on  the five invariants $\calI_1 ... \calI_5$ and must be determined. If $G^{(1)} = -0.09, G^{(n)} =0$, the NLEVM degenerates to the classical $k-\epsilon$ model. When the NLEVM was proposed, it was impossible to find good approximations for these functions  and as a result, it did not receive adequate attention.  
This hurdle however was overcome with the help of machine learning  \cite{ling2016reynolds} where   $G^{(i)}$ were learned from high-fidelity simulation data. Neural networks with parameters $\bt$ were employed for the coefficients i.e. $G_{\bm \theta}^{(i)}$  and: 
\begin{align}\label{eq:TBNN_model}
	\bm{b}_{\bt} = \sum_{i=1}^{10}G_{\bm \theta}^{(i)}(\underbrace{\calI_1 ... \calI_5}_{\substack{Scalar\\Invariants}}) \bm{\mathcal{T}}^{(i)}; \quad \bm{\tau}_{\bm \theta} = 2k \bm{b}_{\bt} + \frac{2k}{3}\bm{I}.
\end{align}
%
We  use $\tauTheta$ to denote the neural-network-based, discretized  RS tensor terms in the subsequent discussions.


As in   \cite{geneva2019quantifying}, we employ the following prior for the NN parameters $\bt$:
\begin{align}
	p(\bm \theta \mid \nu) = \mathcal{N}(\bm \theta | 0, \nu^{-1} \bm I_{d_{\theta}}), \quad p(\nu) = Gamma(\nu|a_0,b_0),
	\label{eq:prior_theta}
\end{align}
where $d_{\theta}=dim(\bt)$ and a Gamma hyperprior was used for the common precision hyperparameter $\nu$ with $(a_0,b_0)=(1.0, 0.02)$. 
The resulting prior has the density of a Student's $\mathcal{T}$-distribution centered at zero, which is obtained by analytically marginalizing over the hyperparameter $\nu$ \cite{bishop2006pattern}. 

\update{Similarly however  to the  most widely used RANS closure models, such as the Launder-Sharma $k-\epsilon$ \cite{Launder1974} or Wilcox's $k-\omega$ \cite{Wilcox2008}, which are based on the Boussinesq turbulent-viscosity hypothesis,   the present model of $\tauTheta$  postulates that the RS tensor at each  point in the problem domain depends on the flow features at the same point (locality assumption). This is a very strong assumption for flows that exhibit strong inhomogeneity \cite{pope2000turbulent}.}
\begin{table}[]
	\centering
	\renewcommand{\arraystretch}{1.3}
	\begin{tabular}{p{0.5cm} p{4.0cm} p{0.5cm} p{4.5cm}}
		\hline
		$\bs{\mathcal{T}}^{(1)}$ &= $\bm{S}$,
		& $\bs{\mathcal{T}}^{(6)}$ &= $\bm{\Omega}^2 \bm{S} + \bm{S} \bm{\Omega}^2 - \frac{2}{3} \mathrm{tr} (\bm{S} \bm{\Omega}^2) \bs{I}$, \\
		$\bs{\mathcal{T}}^{(2)}$ &= $\bm{S} \bm{\Omega} - \bm{\Omega} \bm{S}$,
		& $\bs{\mathcal{T}}^{(7)}$ &= $\bm{\Omega} \bm{S} \bm{\Omega}^2 + \bm{\Omega}^2 \bm{S} \bm{\Omega}$,  \\
		$\bs{\mathcal{T}}^{(3)}$ &= $\bm{S}^2 - \frac{1}{3}\mathrm{tr}(\bm{S}^2) \bs{I}$,
		& $\bs{\mathcal{T}}^{(8)}$ &= $\bm{S} \bm{\Omega} \bm{S}^2 - \bm{S}^2 \bm{\Omega} \bm{S}$, \\
		$\bs{\mathcal{T}}^{(4)}$ &= $\bm{\Omega}^2 - \frac{1}{3}\mathrm{tr}(\bm{\Omega}^2) \bs{I}$,  &
		$\bs{\mathcal{T}}^{(9)}$ &= $\bm{\Omega}^2 \bm{S}^2 + \bm{S}^2 \bm{\Omega}^2 - \frac{2}{3} \mathrm{tr} (\bm{S}^2 \bm{\Omega}^2) \bs{I}$,\\
		$\bs{\mathcal{T}}^{(5)}$ &= $\bm{\Omega} \bm{S}^2 - \bm{S}^2 \bm{\Omega}$,
		& $\bs{\mathcal{T}}^{(10)}$ &= $\bm{\Omega} \bm{S}^2 \bm{\Omega}^2 - \bm{\Omega}^2 \bm{S}^2 \bm{\Omega}$, \\
		\hline
	\end{tabular}
 \caption[Complete set of basis tensors $\bs{\mathcal{T}}^{(n)}$ for a nonlinear eddy viscosity model]{Complete set of basis tensors $\bs{\mathcal{T}}^{(n)}$, that can be formed form $\bm{S}$ and $\bm{\Omega}$. Matrix notation is used for clarity. The trace of a tensor is denoted by $\mathrm{tr}(\bm{S}) = S_{ii}$.}
 \label{table:tensor_basis}
\end{table}

\subsubsection{Stochastic discrepancy term to RS}\label{sec:discrepancy tensor}
We argue that despite the careful selection of input features of the mean velocity field and the flexibility in the resulting map afforded by the NN architecture, the final form might not be able to accurately capture the true RS or at least not at every grid point  in the problem domain. This would be the case, if e.g. non-local terms, which are unaccounted in the aforementioned formulation,  played a significant role.
As mentioned earlier, this gives rise to model uncertainty of an aleatoric nature which is  of a different type than the  epistemic uncertainty in the parameters $\bt$ of the closure model presented in the previous section. It is this model uncertainty that we propose to capture with the latent, random vector $\bm{\epsilon}_{\tau}$ in \refeq{eq:taudecomp}. As for $\bm{\tau}_{\bt}$ in the previous section, $\bm{\epsilon}_{\tau}$ is a vector containing the contribution from all grid points in the problem domain. Hence, in the two-dimensional setting and given the symmetry of the RS tensor, $\bm{\epsilon}_{\tau}$ would be of dimension $d_{\epsilon}=3N$ where $N$ is the total number of grid points.

Before discussing the prior specification for $\epsTau$ and associated inference procedures,  we propose a dimension-reduced representation that would facilitate subsequent tasks given the high values that $N$ takes in most simulations. In particular, we represent $\epsTau$ as:
\be
\epsTau=\bm{W} \bm{E_{\tau}}.
\label{eq:epsdim} 
\ee
It is based on considering   $N_d$  subdomains of the problem domain and  assuming that for all grid points in a certain subdomain, the corresponding RS discrepancy terms are identical. 
This can be expressed as in  \refeq{eq:epsdim} above where the entries of $\bm{W}$ are $1$ if a corresponding grid point (row of $\bm{W}$) belongs in a certain subdomain (column of $\bm{W}$) and $0$ otherwise. \update{A non-binary $\bs{W}$ would also be possible, although in this case its physical interpretation in terms of subdomains would be  occluded.} The vector $\bm{E}_{\tau} =\{ \bm{E}_{\tau,J}\}_{J=1}^{N_d}$ contains therefore the RS discrepancy terms  for each subdomain, e.g. $dim( \bm{E}_{\tau,J})=3$ for a two-dimensional flow.
\update{The dimensionality reduction scheme resembles Principal Component Analysis (PCA)  \cite{tipping1999probabilistic}. In contrast to the latter however, we never have data/observations  of the vectors to be reduced, i.e. $\bs{\epsilon_{\tau}}$ in our case. These are inferred implicitly from the LES data and simultaneously with $\bs{W}$.} In the ensuing numerical illustrations, the division into subdomains is done in a regular manner and $\bm{W}$ is prescribed a priori. One could nevertheless readily envision a learnable $\bm{W}$ or even an adaptive refinement into subdomains. 

The incorporation of the model discrepancy variables $\bm{\epsilon}_{\tau}$ or $\bm{E}_{\tau}$  at each grid-point or subdomain introduces redundancies i.e.  there would be an infinity of combinations of $\bt$ and $\bm{\epsilon}_{\tau}$/$\bm{E}_{\tau}$  that could fit the data equally well. 
In order to address this issue, we invoke the concept of {\em sparsity} which has been employed in similar situations in the context of physical modeling  \cite{brunton_discovering_2016,felsberger_physics-constrained_2019}. To this end, we make use of  a sparsity-enforcing Bayesian  prior based on the Automatic Relevance Determination (ARD)  \cite{neal2012bayesian,10.5555/2968826.2969008}. 
In particular:
\be
p(\bm{E}_{\tau}|\bm{\Lambda})=\prod_{J=1}^{N_d} p(\bm{E}_{\tau, J} | \bm{\Lambda}^{(J)})= \prod_{J=1}^{N_d} \mathcal{N}\left( \bm{E}_{\tau, J}  ~|\bm{0}, ~diag(\bm{\Lambda}_J)^{-1}) \right),
\label{eq:gprior}
\ee
where $\bm{E}_{\tau,J}$ denotes the stochastic RS discrepancy term at subdomain $J$ 
and the vector of hyperparameters  $\bm{\Lambda}_J$ contains the corresponding precisions (e.g.  for two-dimensional flows, $dim(\bm{\Lambda}_J)=3$).  A-priori therefore we assume that the RS discrepancies are zero on average with an unknown variance/precision that will be learned from the data as it will be discussed in the sequel. 
This is combined with the following  hyperprior (omitting the hyperparameters $\alpha_0,\beta_0$):
\be
p(\bm{\Lambda})=\prod_{J=1}^{N_d} \prod_{\ell=1}^L Gamma(\Lambda_{J,\ell} ~| \alpha_0,\beta_0),
\label{eq:lprior}
\ee
where $\Lambda_{J,\ell} $ denotes the $\ell^{th}$ entry (e.g. $L=3$ for two-dimensional flows)  of the vector of precision hyperparameters in subdomain $J$.
We note that when $\Lambda_{J,\ell} \to  \infty$, then the corresponding model discrepancy term $E_{\tau, J,\ell} \to 0$. The resulting prior for $\bm{E}_{\tau}$ arising by marginalizing
the hyperparameters $\bm{\Lambda}$ is a light-tailed, Student’s t-distribution \cite{tipping_relevance_2000} that promotes solutions in the vicinity of $0$ unless strong evidence in the data suggests otherwise. The hyperparameters
$\alpha_0,\beta_0$  are effectively the only ones that need to be provided by the analyst. We advocate very
small values ($\alpha_0=\beta_0=10^{-3}$  was used in the ensuing numerical illustrations) which correspond to an uninformative prior \cite{bishop_variational_1999}. 
\update{This is consistent with the values adopted in the original paper where the ARD prior was employed in conjunction with variational inference i.e. in \cite{bishop_variational_1999}.}

\update{The dimension of the $\epsTau$ is high i.e.  $3N$ in 2D (and $6N$ in 3D) where $N$ is the total number of grid points. In contrast, the dimension of $\bt$ is independent of the grid as the parametric part of the closure model is the same at all grid points. 
Depending on the amount of data, one can envision cases where multiple combinations of $\bs{\epsilon}_{\tau}$ and $\bt$ could fit the data equally well. Non-zero values of $\bs{\epsilon}_{\tau}$ would imply model errors at the corresponding grid points. Ceteris paribus, we would prefer the combination with the least number of model errors or equivalently the one for which our parametric model can capture for most of the domain the closure accurately. It is indeed these types of solutions/combinations that the ARD prior promotes.}

\subsubsection{Data, likelihood  and posterior}\label{sec:model parametrization}



The probabilistic model proposed is trained with {\em indirect} observational data that pertain to time-averaged velocities and pressures at various points in the problem domain. 
This is in contrast  to the majority of efforts in data-driven RANS closure modeling  \cite{geneva2019quantifying,ling2016reynolds,wang2017physics,kaandorp2020data}, which employ {\em direct} RS data. In the ensuing numerical illustrations, the data is obtained from higher-fidelity computational simulations, but one could readily make use of actual, experimental  observations.

In particular, we consider $M \ge 1$ flow settings and denote the observations collected as  $\calD = \{\zobs^{(m)}\}_{m=1}^M$. These  consist of time-averaged velocity/pressure values where $dim(\zobs^{(m)})=N_{obs}$. The locations of these measurements do not  necessarily  coincide with the mesh used to solve the RANS model in  \refeq{eq:RANS_equation} nor is it  necessary that the same number of observations is available for each of the $M$ flow settings. The data is ingested with the help of a Gaussian likelihood:
\be
\begin{array}{ll}
	p( \mathcal{D} \mid \bt,\epsTau^{(1:M)}) & = \prod_{m=1}^M p(\zobs^{(m)} \mid \bt,\epsTau^{(m)} )\\
	& =\prod_{m=1}^M \calN(\zobs^{(m)} \mid \bm{z}( \bt, \epsTau^{(m)}),\bs{\Sigma} ),
\end{array}
\label{eq:likelihood}
\ee
where $\bm{z}( \bt, \epsTau^{(m)})$ denotes the solution vector of the discretized RANS equations (see \refeq{eq:RANS_equation}) with the closure model suggested by \refeq{eq:taudecomp} i.e.  $\bm{\tau}=\bm{\tau}_{\bt}(\bm{u})+ \epsTau^{(m)}$.
We note that a different set of latent  variables $\epsTau^{(m)}$ is needed for each flow scenario as, by its nature,  model discrepancy will in general assume different values under different flow conditions. 
\update{We also note that the $\epsTau$ is grid-dependent or problem geometry-dependent, which restricts it from making predictions for a completely different flow geometry. One could, with appropriate modeling enhancements, learn the dependence of $\bs{\epsilon}_{\tau}$ on flow parameters (e.g. boundary/initial conditions, geometry, $Re$ number) which would make it usable in other flow geometries. Alternatively, it could be trained under different flow settings (e.g. different $Re$ numbers as we do) and learn, on aggregate, the model error which in turn  could be used to make predictions under different flow settings (as in our case for different $Re$ number).}

We denote with $\bs{\Sigma}$ the covariance matrix \update{of the Gaussian likelihood} which,  given the absence of actual observation noise, plays the role of a tolerance parameter determining the tightness of the fit. The covariance was expressed as $\bs{\Sigma} = diag(\sigma_1^2, \cdots, \sigma_{3N_{obs}}^2)$, where the values in the diagonal vector are set to $1\%$ of the mean  of the squares  of each observable across the $M$ flow settings i.e.  $\sigma_i^2=0.01~\frac{1}{M}\sum_{m=1}^M \left( \hat{z}_i^{(m)} \right)^2$. 

\begin{figure}[!htbp]
	\centering
	\includegraphics[width=0.5\textwidth]{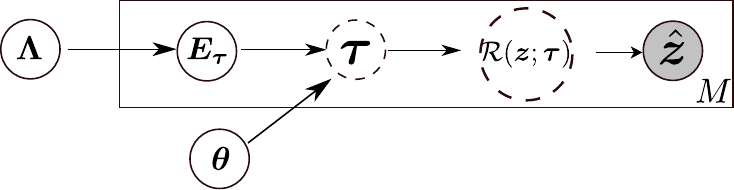}
	\caption{
		Probabilistic graphical model  of  the proposed model including  model parameters ($\bm \theta$, $\bm \Lambda$), latent variables ($\ETau$) and  observables $\zobs$  from $M$  flow scenaria.  Deterministic nodes are indicated with circles with dashed line, stochastic with circles with solid line and known/observed are shaded. 
		}
	\label{fig:probab_graph}
\end{figure}

%
%
By combining the likelihood above with the priors discussed in the previous sections as well as by employing the dimensionality reduction scheme of \refeq{eq:epsdim} according to which  $ \epsTau^{(m)}$ can be expressed as $ \epsTau^{(m)}=\bm{W} \bm{E}_{\tau}^{(m)}$, we obtain a posterior on:
\bi
\item the parameters $\bt$ of the parametric closure model,
\item the latent variables $\bm{E}_{\tau}^{(1:M)}$ expressing the stochastic model discrepancy in {\em each} of the $M$ training conditions,
\item the hyperparameters $\bm{\Lambda}$ in the hyperprior of $\bm{E}_{\tau}^{(1:M)}$,
\ei
which would be of the form (omitting given hyperparameters):
\be
\begin{array}{ll}
	p(\bt, \bm{E}_{\tau}^{(1:M)}, \bm{\Lambda} |~ \mathcal{D})  & \propto p(\mathcal{D}~ |~ \bt, \bm{E}_{\tau}^{(1:M)}) ~p( \bm{E}_{\tau}^{(1:M)} | \bm{\Lambda})~p(\bt)  ~p(\bm{\Lambda}) \\
	& = \left(  \prod_{m=1}^M p(\zobs^{(m)} \mid \bt, \bm{E}_{\tau}^{(m)})~p(\bm{E}_{\tau}^{(m)}| \bm{\Lambda}) \right)~p(\bt)  ~p(\bm{\Lambda}).
\end{array}
\label{eq:post}
\ee
%
%
%
An illustration of the  corresponding graphical model  is contained  in \reffig{fig:probab_graph}.

\subsubsection{Inference and Learning}\label{sec:learning}

On the basis of the probabilistic model proposed and the posterior formulated in the previous section, we discuss numerical inference strategies for identifying the unknown parameters and latent variables.
The intractability of the posterior stems from the likelihood which entails the solution of the discretized RANS equations.
%
%
%
We advocate  the use of Stochastic Variational Inference (SVI) \cite{hoffman_stochastic_nodate} which results in a closed-form approximation of the  posterior $p(\bt, \bm{E}_{\tau}^{(1:M)}, \bm{\Lambda} | \mathcal{D})$. In contrast to the popular, sampling-based strategies (MCMC, SMC etc.), SVI yields biased estimates at the benefit of computational efficiency. 
\update{Readers interested in comparative studies/discussions on SVI and MCMC in terms of accuracy and computational efficiency  are directed to \cite{blei2017variational}.}
Given a family of probability densities  $q_{\bm \xi} \left( \bm \theta,\bm \Lambda,  \bm{E}_{\tau}^{(1:M)} \right) $ parametrized by $\bm \xi$, we find the optimal, i.e. the one that is closest to the exact posterior in terms of their Kullback-Leibler divergence, by maximizing the Evidence Lower Bound (ELBO) $\mathcal{F}(\bm \xi)$ \cite{bishop2006pattern}:
\begin{align}
	\mathcal{F}(\bm \xi) & = \mathbb{E}_{q_{\xi} \left( \bm \theta,\bm \Lambda, \bm{E}_{\tau}^{(1:M)} \right)} \left[
	\log \left( \frac{p \left( \mathcal{D},\bm \theta,\bm \Lambda, \bm{E}_{\tau}^{(1:M)} \right)}{q_{\xi} \left( \bm \theta,\bm \Lambda, \bm{E}_{\tau}^{(1:M)} \right)} \right)
	\right] \nonumber \\
	& =  \mathbb{E}_{q_{\xi} \left( \bm \theta,\bm \Lambda, \bm{E}_{\tau}^{(1:M)} \right)} \left[
	\log \left( 
	\frac{ p(\mathcal{D}~ |~ \bt, \bm{E}_{\tau}^{(1:M)}) ~p( \bm{E}_{\tau}^{(1:M)} | \bm{\Lambda})~p(\bt) ~p(\bm{\Lambda})}
	{ q_{\xi} \left( \bm \theta,\bm \Lambda, \bm{E}_{\tau}^{(1:M)} \right)} 
\right)
	\right].
	\label{eq:ELBO}
\end{align}
As its name suggests, it can be readily shown and that ELBO lower-bounds the model log-evidence and their difference is given by the aforementioned KL-divergence i.e.:
\be
\log p(\mathcal{D})=\mathcal{F}(\bm \xi)+KL \left( q_{\xi} ( \bm \theta,\bm \Lambda, \bm{E}_{\tau}^{(1:M)} )~ \middle| \middle| ~p ( \bm \theta,\bm \Lambda, \bm{E}_{\tau}^{(1:M)}| \mathcal{D} ) \right).
\ee
%
%
%
%
In order to expedite computations we employ a mean-field assumption \cite{blei2017variational} according to which the approximate posterior is factorized as:
\begin{align}
	q_{\bm \xi} \left( \bm \theta,\bm \Lambda, \bm{E}_{\tau}^{(1:M)}\right)  = q_{\bm \xi}(\bm \theta)q_{\bm \xi}(\bm \Lambda)\prod_{m=1}^{M}q_{\bm \xi}\left(\bm{E}_{\tau}^{(m)}\right).
	\label{eq:qmean}
\end{align}
We employ Dirac-deltas for the first two densities i.e.:
%
\be
q_{\bm \xi}(\bm \theta) = \delta(\bm \theta - \thetaMAP),
\label{eq:posttheta}
\ee
\be
q_{\bm \xi}(\bm \Lambda) = \delta(\bm \Lambda - \LambdaMAP).
\label{eq:postlambda}
\ee
In essence, we obtain point estimates for $\bt, \bm{\Lambda}$ which coincide with the  Maximum-A-Posteriori (MAP) estimates. \update{The specific choice of Dirac-deltas was motivated by computational cost reasons in the inference and prediction steps. Furthermore, given that the dimension of $\bt$ is generally (much) smaller than the number of observations (see section \ref{sec:numerical Illustrations}), we anticipate that its posterior would not deviate much from a Dirac-delta. 
The final reason for which posterior uncertainty (albeit small) on $\bt$ was not included, was to emphasize the significance and impact of the stochastic, model correction term that we propose.} 

For the model discrepancy variables  $\bm{E}_{\tau}^{(m)}$, we employ uncorrelated Gaussians given by:
\be
q_{\bm \xi}(\bm{E}_{\tau}^{(m)}) = \calN(\bm{E}_{\tau}^{(m)} \mid \bm{\mu}^{(m)}_{E}, \text{diag}(\bm{\sigma}^{2,(m)}_{E})), \quad \forall i \in \{1,\ldots,M\}.
\label{eq:poste}
\ee

In summary, the vector $\bm \xi$ of the parameters in the variational approximation consists of:
\begin{align}
	\bm{\xi} = \{\thetaMAP, \LambdaMAP, \{\bm{\mu}^{(m)}_{E}, \bm{\sigma}^{2,(m)}_{E}\}_{m=1}^M\}.
\end{align}

The updates of the parameters $\bm \xi$ are carried out using derivatives of the ELBO. These entail expectations with respect to $q_{\bm \xi}$ which are estimated (with noise) by Monte Carlo in conjunction with  the ADAM stochastic optimization scheme \cite{kingma2014adam}. 
In order to reduce the Monte-Carlo noise in the estimates, we employ the reparametrization trick \cite{kingma2013auto}. This is made possible here 
due to the form of the approximate posterior $q_{\bm \xi}$.  In particular, if we summarily denote with $\bm \eta = \{\bm \theta,\bm \Lambda, \bm{E}_{\tau}^{(1:M)}\}$ and given that the approximate posterior $q_{\bm{\xi}}(\bm \eta)$ can be represented by deterministic  transform $\bm \eta = g_{\xi}(\bm \phi)$, where $\bm \phi$ follows a known density $q(\bm \phi)$\footnote{
Based on the form of $q_{\bm{\xi}}$ in Equations (\ref{eq:qmean}),(\ref{eq:posttheta}), (\ref{eq:postlambda}), (\ref{eq:poste}), the transform employed can be written as $\bm \theta=\bm \theta_{MAP}$, $\bm \Lambda=\bm \Lambda_{MAP}$ and $\bm{E}_{\tau}^{(m)} =  \bm{\mu}^{(m)}_{E}+ \text{diag}(\bm{\sigma}^{2,(m)}_{E}) ~\bm \phi$ where $q(\bm \phi)=\mathcal{N}(\bm \phi | ~\bm 0, \bm I)$.
}, the expectations involved in the ELBO  and, more importantly, in its gradient  can be rewritten as:
\begin{align}
	\nabla_{\bm \xi} \calF(\bm \xi) & = \mathbb{E}_{q(\bm \phi)} \left[ 
	\nabla_{\bm{\xi}}g_{\xi}(\bm \phi)\nabla_{\bm \eta}
	\left(\log  p ( \mathcal{D},\bm{\eta})- 
	\log q_{\xi} (\bm{\eta})\right)   \right].
	\label{eq:ELBO_grad_est_cont}
\end{align}
One observes that derivatives of the log-likelihood with respect to $\bm \eta$ are needed. This in turn would imply derivatives of the RANS-model outputs with respect to $\{\bm \theta,\bm \Lambda, \bm{E}_{\tau}^{(1:M)}\}$ which appear indirectly through $\bm \tau$. Such derivatives are rendered possible by using an adjoint formulation of the discretized RANS model that yields in effect a {\em  differentiable} solver. 

%
%

Further details about the derivatives of the ELBO and the use of RANS-model sensitivities obtained by an adjoint formulation can be found in \ref{app:grad ELBO}. 
An algorithmic summary of the steps entailed  is contained  in Algorithm \ref{Alg:SVI_algo}. 
\update{We note finally that the ELBO $\mathcal{F}$ (which serves as a lower bound to the model evidence) and  can be used to compare (in a Bayesian sense) alternative models. Such model variations could arise by changing e.g. $\bs{W}$ i.e. the partition into subdomains which enables the dimensionality reduction of the stochastic model discrepancy vector. One could even employ   an adaptive  division into subdomains guided by $\mathcal{F}$ so as the data can dictate which regions require more/less refinement.}%

\begin{algorithm}[t]
\SetAlgoNlRelativeSize{0} 
\begin{spacing}{1.0}  
	\SetAlgoLined
	\SetKwData{Data}{Data}
	\SetKwInOut{Input}{Input}
	\SetKwInOut{Output}{Output}
	\Input{$\calD = \{\zobs^{(i)}\}_{m=1}^M$, $\bm{W}$}
	\Output{$\bm{\xi}=\{\thetaMAP, \LambdaMAP, \{\bm{\mu}^{(m)}_{E}, \bm{\sigma}^{2,(m)}_{E}\}_{m=1}^M\}$}
	\While{ELBO $\hat{\mathcal{F}}$ not converged}{
	    $\bt \leftarrow \bt_{MAP}$, ~~
	    $\bm{\Lambda} \gets \bm{\Lambda}_{MAP}$ \;
		\vspace{2mm}
		\For{$m \in \{1:M\}$}
{
		\For{$k \in \{1:K\}$}
{		
 \tcp{Reparametrization trick}
		Sample $\bm{\phi}^{(m,k)} \sim \mathcal{N}(~\bm{0,~I})$  for $k = 1, \cdots, K$ \;		
		\tcp{Compute stochastic discrepancy terms}
		$\ETau^{(m,k)}=g_{\bm{\xi}}(\bm{\phi}^{(m,k)})=\bm{\mu}^{(m)}_{E}+  \bm{\sigma}^{(m)}_{E}\odot \bm{\phi}^{(m,k)}$ \; 
		\vspace{2mm}
		\tcp{Solve  discretized RANS equations}
		Solve $\mathcal{R}(\z; \taub_{\bt}(\bm{u})+\bm{W}\ETau^{(m,k)}) = 0$ to obtain the solution vector $\z^{(m,k)}$ \tcp*[r]{\refeq{eq:residual_tau_theta}}
		\vspace{2mm}
		\tcp{Compute log-likelihoods and their gradients}
		$\ell^{(m,k)}(\bt,\ETau^{(m,k)}), ~\pa \ell^{(m,k)}/\pa \bt, ~\pa \ell^{(m,k)}/\pa \ETau^{(m,k)} $ 
		\tcp*[r]{Equations (\ref{eq:loglikem}), (\ref{eq:grad_wrt_theta}), (\ref{eq:grad_wrt_ETau})  }
}
}

\tcp{Monte Carlo estimate of ELBO}
			Estimate ELBO $\mathcal{F}$  using \refeq{eq:elbomc} \;
		\vspace{2mm}
\tcp{Monte Carlo estimate of the gradient of the ELBO}
		Estimate gradient of ELBO $\nabla_{\bm{\xi}} \mathcal{F} $ using \refeq{eq:elbodermc} \;
		\vspace{2mm}
		\tcp{Stochastic Gradient Ascent}
		$\bm{\xi}^{(n+1)} \leftarrow \bm{\xi}^{(n)} + \bm{\rho}^{(n)} \odot \nabla_{\bm{\xi}} \mathcal{F} $ \;
		$n \leftarrow n + 1$ \;
	}
	\Return{$\bm{\xi}$ } 
	\caption{Inference and Learning  using SVI }
	\label{Alg:SVI_algo}
\end{spacing}
\end{algorithm}

\subsubsection{Predictions}\label{sec:prediction}
%


In this section, we describe  how {\em probabilistic}, predictive estimates  of any quantity of interest related to the RANS-simulated flow can be produced using the trained model.
In particular, one can obtain a {\em predictive, posterior density} $p(\bm{z}| \mathcal{D})$ on the whole solution vector  $\bm{z}$ of the RANS equations as follows:
\be
\begin{array}{ll}
	p(\bm{z} | \mathcal{D}) & = \int p(\bm{z}, \bm{E}_{\tau}, \bt, \bm{\Lambda}|\mathcal{D})~d\bm{E}_{\tau}~d\bt~d\bm{\Lambda} \\
	& = \int  p(\bm{z} |\bm{E}_{\tau}, \bt)~p(\bm{E}_{\tau}, \bt , \bm{\Lambda} | \mathcal{D})~d\bm{E}_{\tau}~d\bt~d\bm{\Lambda} \\
	& =\int  p(\bm{z} |\bm{E}_{\tau}, \bt)~p(\bm{E}_{\tau} | \bm{\Lambda}) p(\bt, \bm{\Lambda}|   \mathcal{D})~d\bm{E}_{\tau}~d\bt~d\bm{\Lambda}. \\
\end{array}
\label{eq:predpost}
\ee
The third of the densities in the integrand  is the  posterior which is substituted by its variational approximation i.e.  $q_{\bm \xi}$ in \refeq{eq:qmean} and for the optimal parameter values $\bm \xi$ identified as described in the previous section. The second of the densities represents the prior model prescribed in \refeq{eq:gprior}. Finally, the first of the densities is simply a Dirac-delta which  \update{ corresponds to the solution of the RANS equations obtained when using the proposed  closure model for given $\bm{E}_{\tau}, \bt$.  Since the RANS solver is a black-box, it would not be possible to propagate the uncertainty in $\bm{E}_{\tau}$ (and potentially $\bt$) otherwise. 
}
In practical terms and given the intractability of this integral, the equation above suggests a Monte Carlo scheme for obtaining samples from $p(\z| \bm{D})$ which involves the following steps. For each sample:
\bi
\item Set $\bt=\bt_{MAP}$, $\bm{\Lambda}=\bm{\Lambda}_{MAP}$. (If a different variational approximation to the posterior $q_{\bm \xi}$  than the one in Equations (\ref{eq:posttheta}), (\ref{eq:postlambda}) were used, then $\bt, \bm{\Lambda}$ would need to be sampled from it).
\item Sample $\bm{E}_{\tau}$ from $p(\bm{E}_{\tau} | \bm{\Lambda}_{MAP}) $ in \refeq{eq:gprior} and compute model discrepancy vector $\bm{\epsilon}_{\tau}=\bm{W} \bm{E}_{\tau} $.
\item Solve the discretized RANS model in \refeq{eq:residual_tau_theta} for $\bm{\tau}=\bm{\tau}_{\bt}(\bm{u})+ \bm{\epsilon}_{\tau}$.
\ei
\update{Other numerical integration techniques (e.g.  Quasi Monte Carlo or Importance Sampling) could be used in order to obtain estimates of the integral with fewer RANS solves.}
The aforementioned steps would need to be repeated for as many samples as desired. Subsequently, the samples can be used to compute statistics of the predictive estimate (e.g. predictive mean, variance, credible intervals, etc) not only of $\bm{z}$ (i.e. velocities/pressures) but of any quantity of interest such as the lift, drag, skin friction, etc. 

We  note however that stochastic RS discrepancy terms $\epsTau$ or $\bm{E}_{\tau}$ and the associated probabilistic model, are limited to the flow geometry used for the training. While  it can be used for unseen flow scenarios (e.g. different $Re$ number, inlet conditions, boundary conditions), it cannot   be employed for a different flow geometry. In theory, the parametrization of the $\epsTau$ can be updated to accommodate different geometries, but we leave it for future investigations. Finally,  we would like to highlight the fact that baseline RANS data {\em is not needed} as an input to the neural networks for prediction in the proposed scheme, as opposed to other frameworks that have been employed in the past (e.g. \cite{geneva2019quantifying,ling2016reynolds,kaandorp2020data,wang2017physics}). 

\begin{figure}[!t]
	\centering
	\includegraphics[width=1\textwidth]{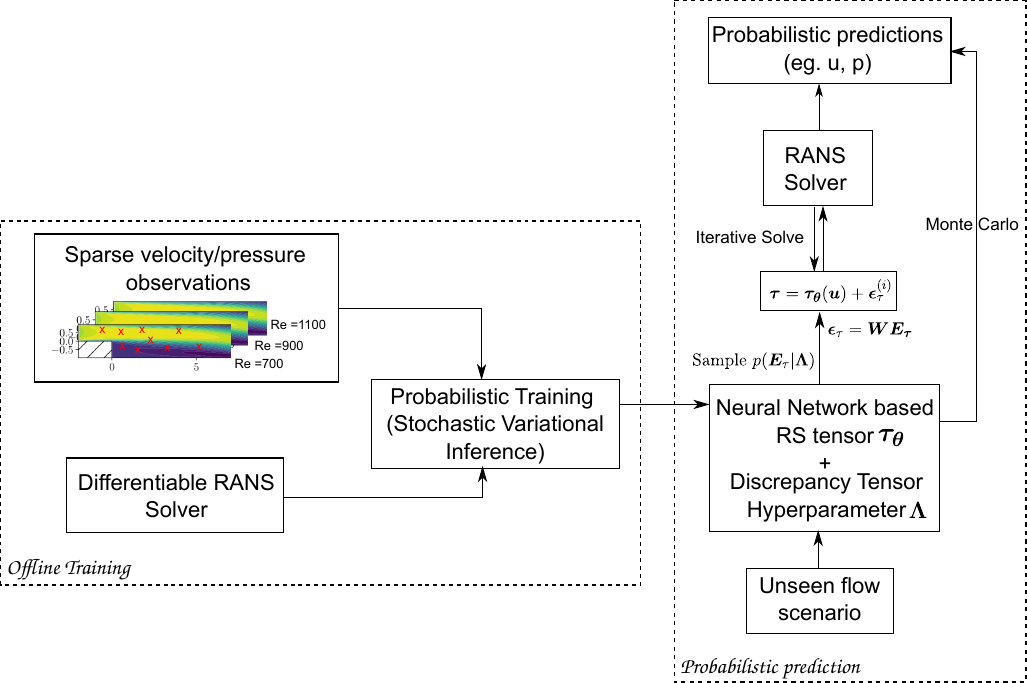}
	\caption{\update{Schematic illustration of the training/inference (left block) and  probabilistic prediction (right block) framework proposed. }}
	\label{fig:flowchart}
\end{figure}

In terms of computational aspects, the stochastic nature of the (reduced) discrepancy tensor $\ETau$ can potentially introduce non-smoothness in the RS vector $\taub$ used for solving the RANS equations. In the present work, we have used diagonal covariance for the hyper-prior of $\ETau$ given by $\bm \Lambda = diag(\bm{\Lambda}^{(J)}),~~ J=1,\ldots,N_d$, thus assuming there is no correlation among the nearby nodes/region. As an avenue for future work, a banded covariance matrix can be employed to capture such spatial correlations. 
Sparsity-inducing priors that account for spatial correlations have been proposed in \cite{Bardsley2013,10.5555/2968826.2969008}. \update{A schematic overview of the methodological framework  discussed in \refsec{sec:learning} and \refsec{sec:prediction} is presented in \reffig{fig:flowchart}}. Numerical  results, training data and the corresponding source code will be made available at \href{https://github.com/pkmtum/D3C-UQ/}{https://github.com/pkmtum/D3C-UQ/} upon publication.

\section{Numerical Illustrations}
\label{sec:numerical Illustrations}
\subsection{Test case: Backward Facing step}

\begin{figure}[!t]
	\centering
	\includegraphics[width=0.75\textwidth]{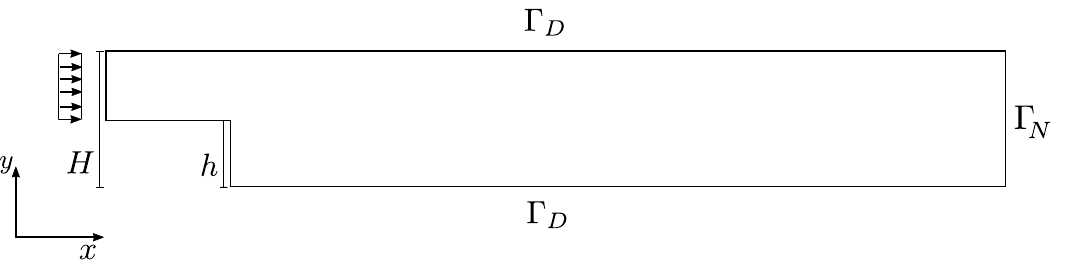}
	\caption{Backward facing step flow configuration with step height $h$ and the total channel height $H$. The origin of the $x-y$ plane is placed at the corner of the step. The axes are depicted at the bottom left to avoid clutter.}
	\label{fig:BFS_flow_geometry}
\end{figure}

We select the backward facing step configuration in order to assess the proposed modeling framework. This is a classic benchmark problem   that has been widely used for studying the performance of turbulence models  as it poses significant challenges due to the complex flow features such as flow separation, reattachment and recirculation \cite{nadge2014high, pioch_turbulence_2023}.
In this setup, as illustrated in \reffig{fig:BFS_flow_geometry} a two-dimensional channel flow is abruptly expanded into a rectangular cavity with a step change in height. The flow separates at the step and forms two recirculation zones downstream, one directly after the step and the other on the upper channel wall downstream. These two recirculation zones affect the reattachment length. 
The flow features can be seen in more detail as depicted in the LES simulations in \reffig{fig:LES_mean_flow}.
In this setting,  the Reynolds number is defined as:
\begin{align}
	Re = \frac{u h}{\nu},
\end{align}
where $h$  
and $\nu$ are the characteristic length (also the step height) and kinematic viscosity, respectively and $u$ denotes the average velocity of the inlet flow. In the present study, the expansion ratio $H/h$ is 2 and the boundary conditions are shown in \reffig{fig:BFS_flow_geometry}. They consist of constant inlet bulk velocity $u_b=1$ in the $x$-direction $(u_b=u)$  on the left boundary, no-slip condition on $\Gamma_D$ (i.e, top/bottom boundary) and zero  tractions along  $\Gamma_N$ (i.e, at the outflow boundary) i.e.  $\left(-p\bm{I}+  \frac{\nu}{2}(\nabla \bm u + \nabla\bm{u}^T) \right)\cdot \hat{n} = 0$ where $\hat{n}$ is the outward normal of the outflow. On the $x-y$ plane, we place the origin at the corner of the step. 
\subsection{Generation of training data}\label{sec:training_data}


In order to generate the training data, we performed Large Eddy Simulations (LES) for various Reynolds numbers i.e. by varying the kinematic viscosity $\nu$. 
 A three-dimensional configuration is adopted wherein the $z$-direction (i.e, the in-plane direction) is periodic and the mean fields averaged over the $z$-direction are used for training.  We also performed RANS simulations using the $k-\epsilon$ model for the same set  of $Re$  numbers to provide a comparison as it is the most widely used RANS model in  industrial applications. In the subsequent discussions, we will refer to the \textbf{$k-\epsilon$ model as the $\emph{baseline}$ RANS model}.
 
We used the open-source CFD platform OpenFOAM \cite{jasak2007openfoam} for the LES and  baseline RANS simulations. 
We utilized the steady-state, incompressible solver simpleFoam for the baseline RANS simulations. This solver uses the Semi-Implicit Method for Pressure Linked Equations (SIMPLE) in order to solve both the momentum and pressure equations.
For the LES simulations, we employed the pimpleFoam transient solver, which combines both the PISO (Pressure Implicit with Split Operator) and SIMPLE algorithms to solve the pressure and momentum equations. In particular, we use the WALE (Wall-Adapting Local Eddy-Viscosity) model \cite{nicoud1999subgrid}. This model is well-suited for capturing turbulent structures near solid walls and is known for its accurate predictions of wall-bounded turbulent flows. \update{In order to overcome the computational demands of LES, a domain decomposition approach was employed. In particular, we split the domain into 16 subdomains and leverage the CPU cores in parallel\footnote{Computations were carried out in a Intel  Core i9-12900K CPU.}}. We discretized both the baseline RANS and LES domains using second-order methods and all the meshes were non-uniform with mesh density increasing in the domains of interest. To ensure numerical accuracy, we ran all simulations with a CFL number below 0.3. Other pertinent details of the LES and baseline simulations  are provided in Table \ref{table:CFD details}. \update{We note that the purpose of the LES model in the present work is to serve as the reference solver that the trained RANS model would try to approximate or match. The framework proposed does not make use of the particulars of the LES solver and one could readily use actual or DNS data (or combinations thereof) for training.  The LES data might deviate from physical reality or the results obtained by DNS. Readers interested in the relative accuracy of LES simulations are directed to \cite{engelmann2021towards,celik2005index}.}

\begin{figure}[!t]
        \centering
        \includegraphics[width=0.95\textwidth,height=.12\textheight]{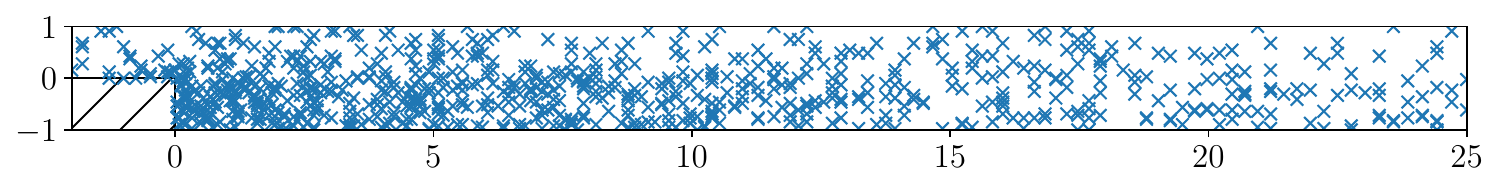}
       \caption{\update{(Random) grid points where LES velocities/pressure were  used for training data. We note that the total number of  LES grid points  is $\mathcal{O}(10^5)$ whereas  LES simulation data at approximately $1000$ grid points were employed.}}
        \label{fig:training_points}
    \end{figure}

\begin{figure}[!t]
	\centering
	\includegraphics[width=0.9\textwidth]{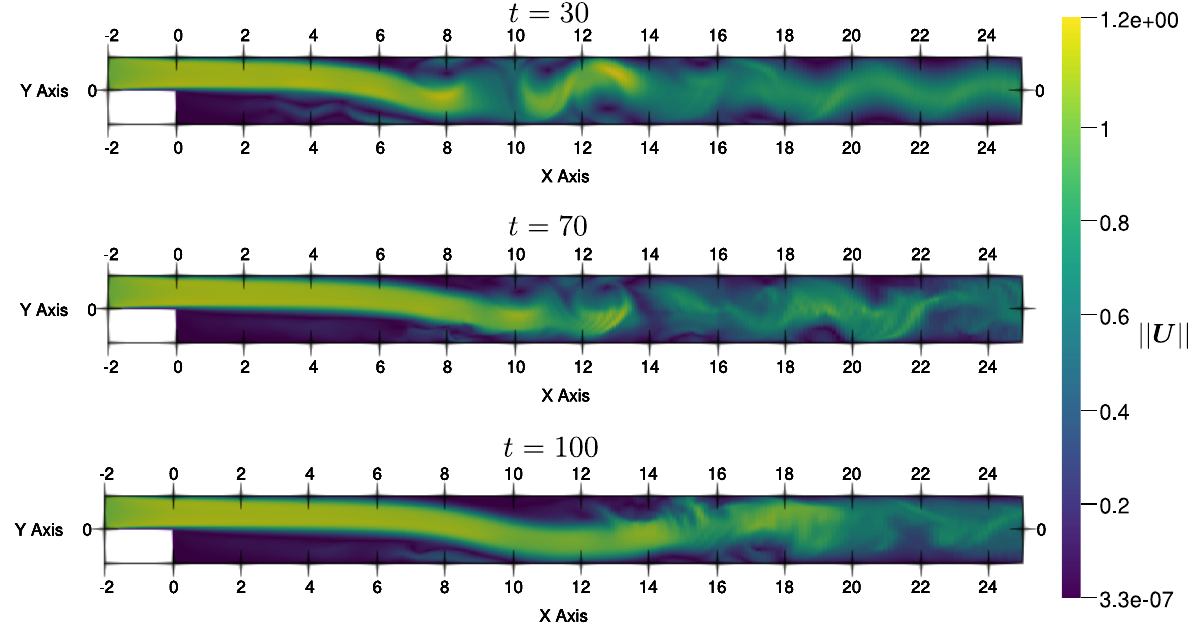}
	\caption{Instantaneous velocity magnitude $||\bm{U}||$ at different time-instants $t=\{30,70,100\}$ obtained from the  LES simulation performed at $Re=1100$. We note that the flow eventually reaches  a stationary state.} 
	\label{fig:LES_snapshots_u}
\end{figure}

\begin{figure}[!t]
	\centering
	\includegraphics[width=0.95\textwidth]{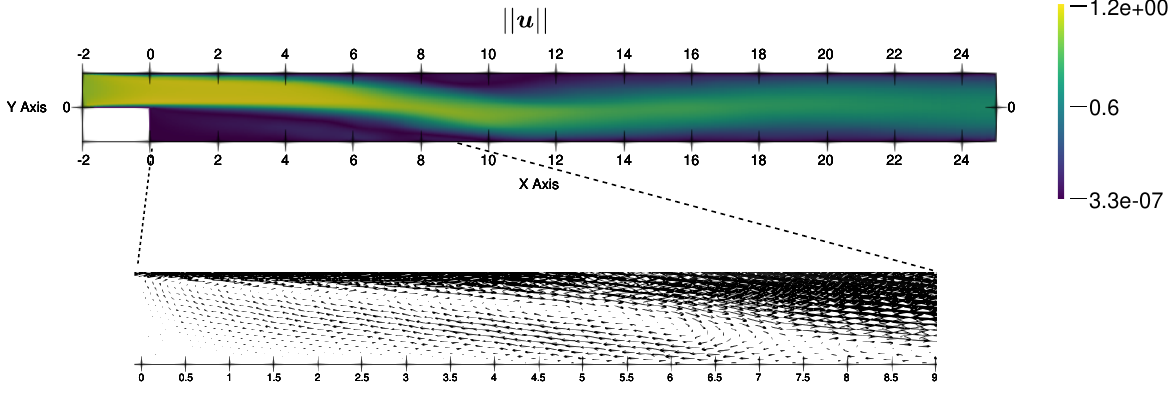}
	\caption{Time-averaged velocity magnitude $||\bm{u}||$ obtained from  LES simulation performed at $Re=1100$. The velocity vector plot highlights the flow separation, recirculation zones and the flow reattachment. 
	}
	\label{fig:LES_mean_flow}
\end{figure}

The mean field reference data from LES is interpolated to the same mesh used for the Finite Element (FE)-based calibrated RANS model (model implementation details discussed in the sequel). However, not all the observations in the grid were used for training. \update{At each RANS grid point, an independent, random coin was flipped with probability $10\%$ of picking the grid point. This resulted in  \textbf{mean velocity/pressure observations at approximately $8\%$ of the grid points to be used as training data}}, i.e. a rather sparse dataset (\update{the selected observation points are depicted in  \reffig{fig:training_points}}). \update{  Naturally, the density of RANS grid was higher near the step, so the density of  observations was higher in the regions of steeper velocity gradients (in the reattachment and recirculation regions) and lower in other regions. Hence, the training dataset $\calD = \{\zobs^{(i)}\}_{m=1}^M$ consists of the velocities/pressure at those grid points obtained from the $M=4$  LES simulations carried out with for the corresponding  $Re$ numbers (in Table 2).  }. The influence of the number of observation points on the learning and predictive results is  an interesting research direction, which we will address in  future investigations. A few snapshots of the instantaneous velocity magnitude for $Re=1100$ from the LES simulation are depicted  in \reffig{fig:LES_snapshots_u} where one  can  observe how the flow features of interest evolve over time. Only the time-averaged velocities/pressures were used for training which are shown  in  \reffig{fig:LES_mean_flow}.

\begin{table}[!htbp]
	\centering
        \begin{tabular}{p{7cm} l}
		\hline
		Domain Size & $27h \times 2h \times h$ \\ 
		DoF LES & 1428920 \\
		DoF $k-\epsilon$ & 35709 \\
		step height ($h$) & 1 \\
		Total channel height ($H$) & 2 \\
			kinematic viscosity $\nu$  values for training data generation ($\times 10^{-3}$)  & $3.33, ~2.00, ~1.42, ~1.11,~0.909$\\
					$Re$ values  for training data generation& $300,700,900,1100$ \\
		$Re$ value for prediction & $500$ \\
		characteristic length & step height $h$\\
		\hline
	\end{tabular}
	\caption{Parameters used for performing the CFD simulations for the generation of the training and test data \cite{gresho1993steady}}
	\label{table:CFD details}
\end{table}

\subsection{Differentiable forward RANS solver and probabilistic learning implementation}
%

For the discretization of the RANS equations (\refeq{eq:residual_tau_theta})  the Finite Element (FE) method was employed and the implementation was carried out in the open source package \texttt{FEniCS} \cite{alnaes2015fenics} due  to its innate  adjoint solver \cite{mitusch2019dolfin}. 
%
The basic quantities and their dimensions  are listed in Table \ref{table:solver details}. Further details about the differentiable solver can be found in \ref{app:RANS solver details}.

Probabilistic inference and learning  tasks were performed using the probabilistic programming package \texttt{Pyro} \cite{bingham2019pyro} which is built on top of the popular machine learning library \texttt{PyTorch} \cite{paszke2019pytorch}. The ELBO maximization was performed using the ADAM scheme \cite{kingma2014adam}. The number of Monte Carlo samples used at each iteration  for the estimation of the ELBO and its gradient was  $K = 5$ (Algorithm \ref{Alg:SVI_algo}). The gradient computation of the ELBO (\refeq{eq:ELBO_grad_est_cont}) was enabled by overloading the \texttt{autograd} functionality of PyTorch to facilitate interaction between the solver's  adjoint formulation and the auto-differentiation-based neural network gradient.
A relatively small learning rate of  $10^{-6}$  was employed due to the Monte Carlo noise in the ELBO gradients.
The neural network architecture employed for the parametric RS model was identical to the one  suggested  by \cite{ling2016reynolds} where the optimal number of hidden layers and nodes per layer was determined to be 8 and 30 respectively. The Leaky ReLU was chosen as the activation function for all  layers. 
We noted however that the usual practice of random weight initialization was unsuitable as it led to divergence of the solution even after applying the stabilization schemes. For this reason, we used baseline RANS closure data  with added noise to pre-train the neural network in order to provide a suitable initialization.

\begin{table}[!htbp]
	\centering
	\begin{tabular}{l l}
		\hline\\ [-1.5ex]
		Quantity & dimensions\\[0.5ex]
		\hline\\ [-1.5ex]
		Domain Size & $27h \times 2h$ \\ 
		number of nodes in FE simulation ($N$) & 12180\\
		dim($\bs{z}$) & 12180 $\times$ 3\\
		dim($\taub$) & 12180 $\times$ 3 \\
		Boolean Matrix dim($\bm{W}$) & 12180 $\times$ 52 \\
		dim($\bm{E}$) & 52 $\times$ 3\\
		dim($\bm{\Lambda}$) & 52 $\times$ 3 \\
		dim($\bm{\theta}$) & 6970 \\[0.5ex]
		\hline
	\end{tabular}
	\caption{Basic quantities and dimensions}
	\label{table:solver details}
\end{table}



\subsection{Results and Discussion}

We assessed the trained model for the test-case with $Re = 500$ which was not contained in the training data. 
In the sequel,  various aspects of the probabilistic predictions obtained as described  in \refsec{sec:prediction} are compared with the reference LES and the baseline RANS predictions. Even though the same {\em parametrized} RS closure term was used in  \cite{ling2016reynolds} (and other subsequent works branching from this), their results are not directly comparable due to the use of blending functions \cite{kaandorp2020data,geneva2019quantifying}, which combine baseline RS values near the walls with  the constant, predicted RS in the bulk and with the amount of blending being case-dependent.

\begin{figure}[!htbp]
\centering
\begin{subfigure}{\textwidth}
\centering
\includegraphics[width=0.85\textwidth]{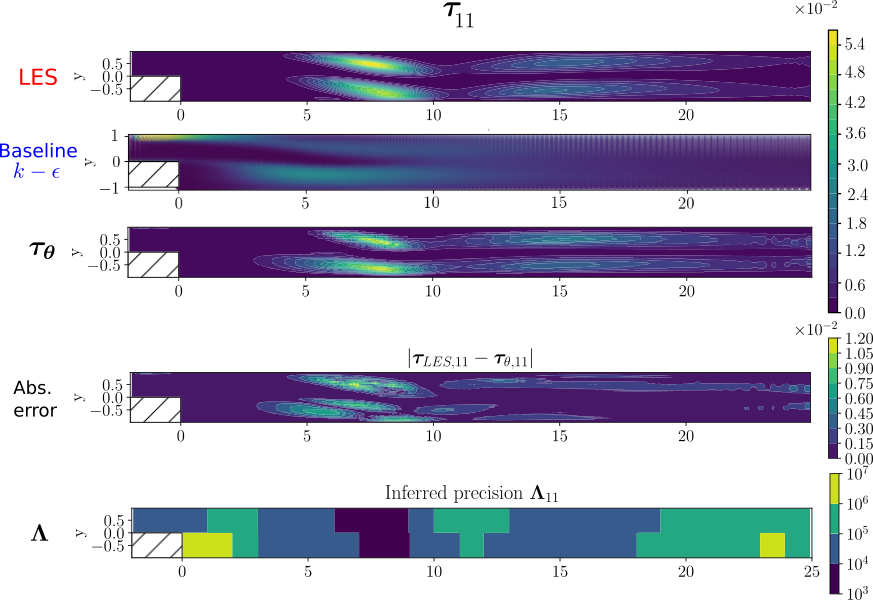}
\caption{Comparison for $\taub_{11}$.}
\end{subfigure}

\bigskip

\begin{subfigure}{\textwidth}
\centering
\includegraphics[width=0.85\textwidth]{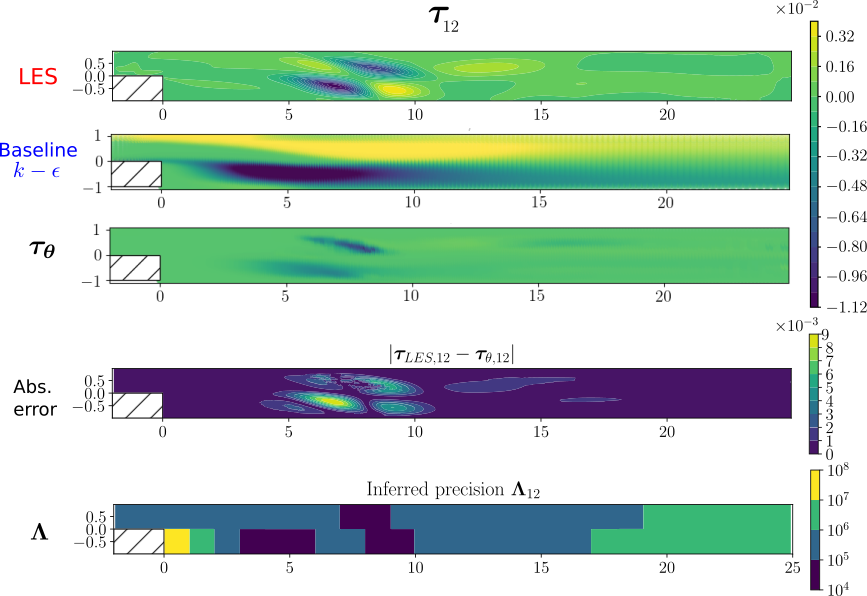}
\caption{Comparison for $\taub_{12}$.}
\end{subfigure}

\caption{\update{Comparison of  the predicted components of the RS tensor $\tauTheta$ with the LES reference values and the \emph{baseline} RANS ($k-\epsilon$) for the test case with $Re = 500$. The three components $\taub_{11}$, $\taub_{12}$ and $\taub_{22}$ are separately compared in subfigures (a), (b), and (c) respectively (\textit{subfigure (c) in the next page}). In each block, \textbf{top} - the LES RS tensor component contour, \textbf{second} - the $k-\epsilon$ RS tensor component contour, \textbf{third} - the predicted, parametric RS tensor $\tauTheta$ for $\bt=\bt_{MAP}$, \textbf{fourth} - the contour plot of the absolute error between the LES RS tensor and the $\tauTheta$ tensor component, \textbf{bottom} -  the inferred hyper-parameter $\bm{\Lambda}$ of the (reduced) discrepancy tensor $\bm{E_{\tau}}$.  }}
\end{figure}
\begin{figure}[ht]\ContinuedFloat
\centering

\begin{subfigure}{\textwidth}
\centering
\includegraphics[width=0.85\textwidth]{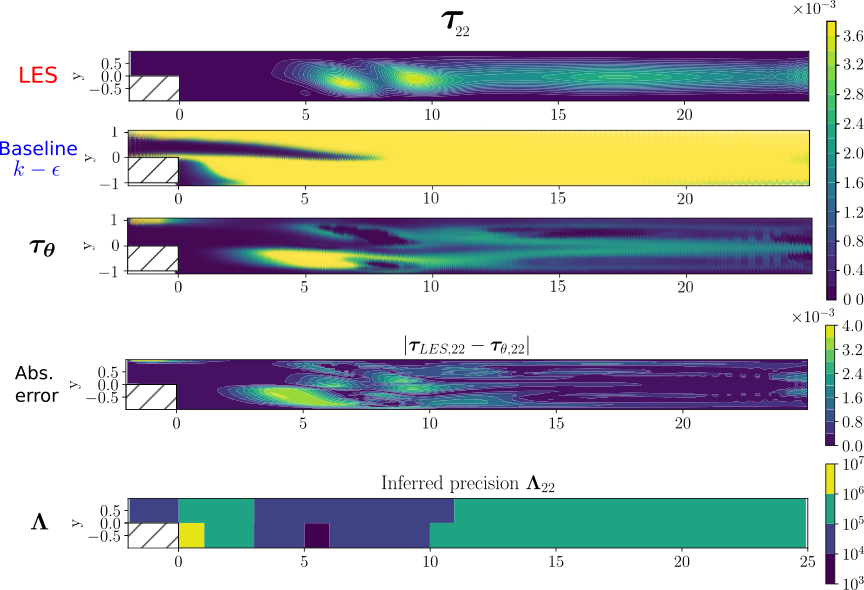}
\caption{Comparison for $\taub_{22}$.}
\label{fig:tau_12}
\end{subfigure}

\caption{\update{(\textit{cont. from the previous page}) Comparison of  the predicted components of the RS tensor $\tauTheta$ with the LES reference values and the \emph{baseline} RANS ($k-\epsilon$) for the test case with $Re = 500$. The three components $\taub_{11}$, $\taub_{12}$ and $\taub_{22}$ are separately compared in subfigures (a), (b) and (c) respectively (\textit{subfigure (a) and (b) in the next page}). In each block, \textbf{top} - the LES RS tensor component contour, \textbf{second} - the $k-\epsilon$ RS tensor component contour, \textbf{third} - the predicted, parametric RS tensor $\tauTheta$ for $\bt=\bt_{MAP}$, \textbf{fourth} - the contour plot of the absolute error between the LES RS tensor and the $\tauTheta$ tensor component, \textbf{bottom} -  the inferred hyper-parameter $\bm{\Lambda}$ of the (reduced) discrepancy tensor $\bm{E_{\tau}}$.  }}
\label{fig:predicted tau comparision}
\end{figure}

The performance of the proposed method in predicting the components of the RS tensor is shown in \reffig{fig:predicted tau comparision}, from which the following conclusions can be drawn:
\begin{itemize}
	\item Even though \textbf{no RS observations} were provided during the training, the parametric $\bs{\tau}_{\bt}$ (i.e.  neural-network based) part of the RS closure is able to capture the basic features of the reference (i.e. LES) RS field. In contrast, the baseline $k-\epsilon$ model severely under/over-estimates its magnitude and misrepresents its spatial variability. \update{We observe that some regions have relatively higher errors (for e.g., around $5< x/h <10$ in \reffig{fig:tau_12} ), which is attributed to the parametric model's inability to provide adequate closure. This model error is exactly what we attempt to capture with our proposed stochastic discrepancy term.}
%

	\item The bottom row of all the three subfigures in \reffig{fig:predicted tau comparision} depicts the inferred precision $\bm \Lambda$ of the (reduced) discrepancy tensor $\ETau$. As this is inversely proportional to the predictive uncertainty we note that it attains smaller values in the regions where the parametric model $\bs{\tau}_{\bt}$ deviates the most from the LES values (e.g. for $5< x/h <10$). Conversely, it attains very large values (which correspond to practically zero model discrepancies) in areas where the parametric closure model is able to correctly account for the underlying phenomena (e.g. far downstream and for all three RS components). 
This is expected as the flow attains an almost  parabolic profile in this region, which in turn translates to reduced fluctuations in the RS tensor, making it easier for the neural network to learn.
\end{itemize}

\begin{figure}[!t]
	\centering
	\includegraphics[width=0.95\textwidth]{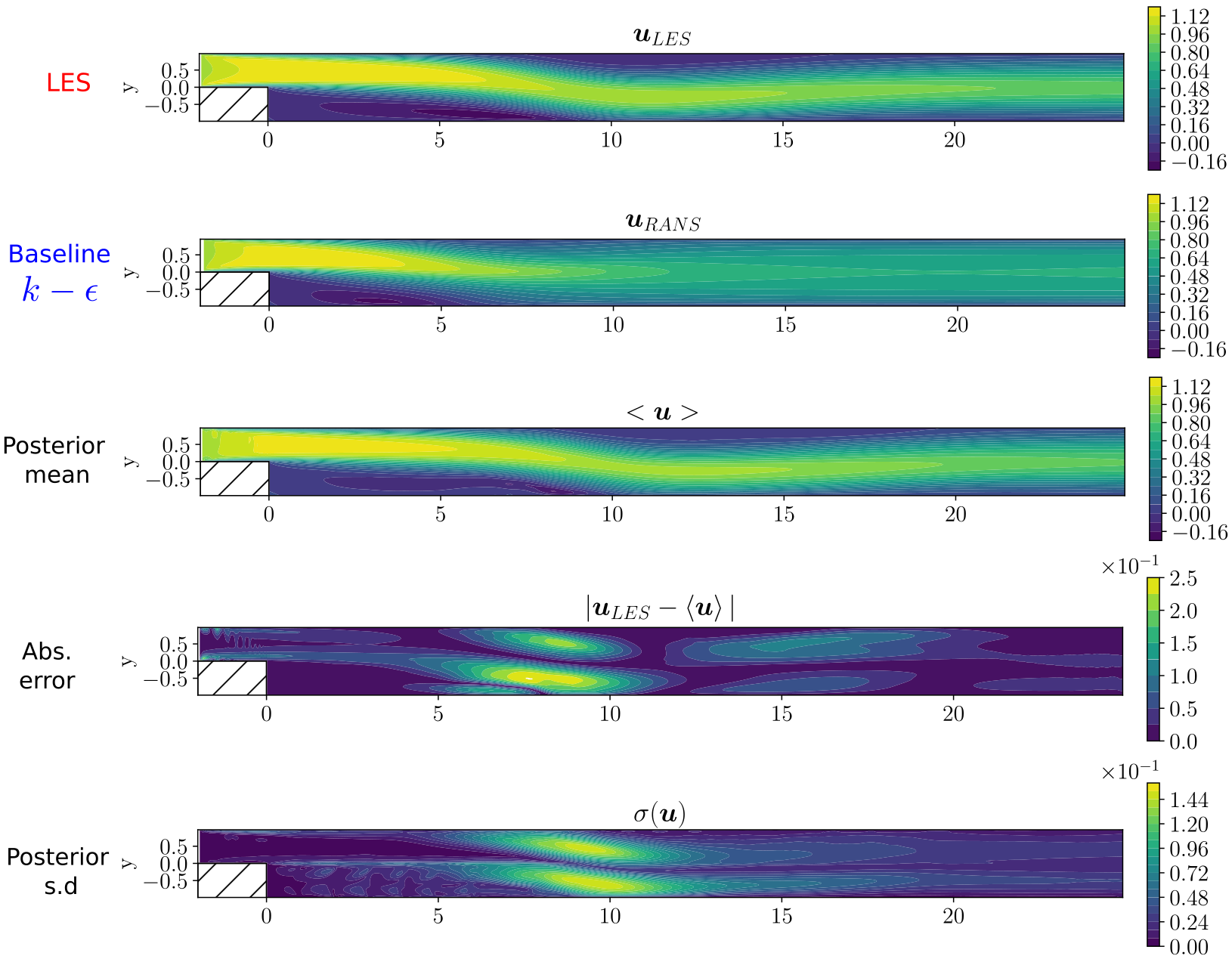}
	\caption{Velocity contours in x-direction ($\bm{u}$) for the test case with Re = 500. \textbf{top} - ground truth (LES), \textbf{second} - baseline $k-\epsilon$, \textbf{third} - the posterior predictive mean ($\left<\bm{u}\right>$), \textbf{fourth} - the contour plot of the absolute error between the LES ($\bm{u}_{LES}$) and the posterior predictive mean ($\left<\bm{u}\right>$), \textbf{bottom} - the standard deviation of the posterior predictive ($\sigma(\bm{u})$).}
	\label{fig:predicted u}
\end{figure}
\begin{figure}[!t]
	\centering
	\includegraphics[width=0.95\textwidth]{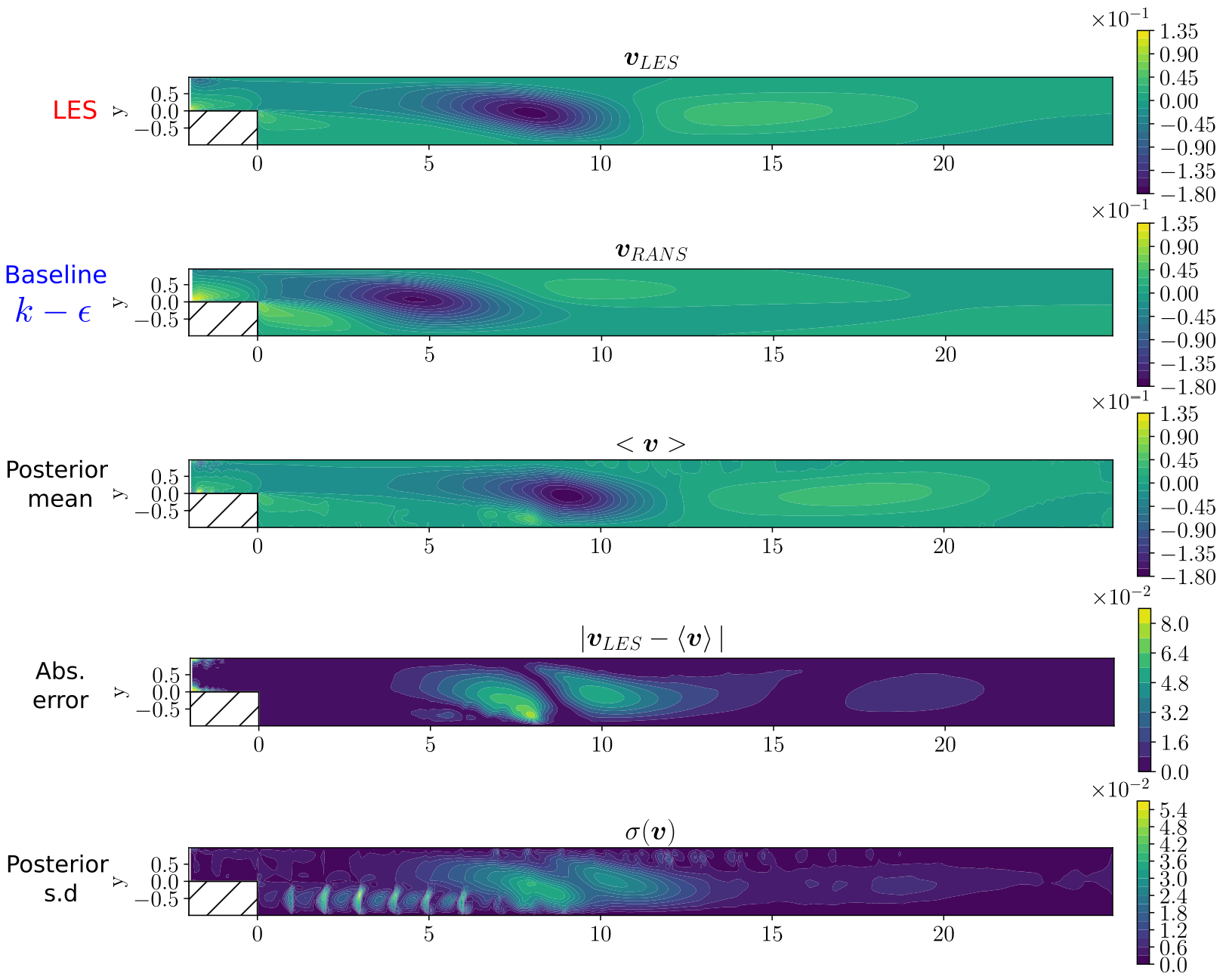}
	\caption{Velocity contours in y-direction ($\bm{v}$) for the test case with Re = 500. \textbf{top} - ground truth (LES), \textbf{second} - baseline $k-\epsilon$, \textbf{third} - the posterior predictive mean ($\left<\bm{v}\right>$), \textbf{fourth} - the contour plot of the absolute error between the LES ($\bm{v}_{LES}$) and the posterior predictive mean ($\left<\bm{v}\right>$), \textbf{bottom} - the standard deviation of the posterior predictive ($\sigma(\bm{v})$).} 
	\label{fig:predicted v}
\end{figure}
\begin{figure}[!t]
	\centering
	\includegraphics[width=0.95\textwidth]{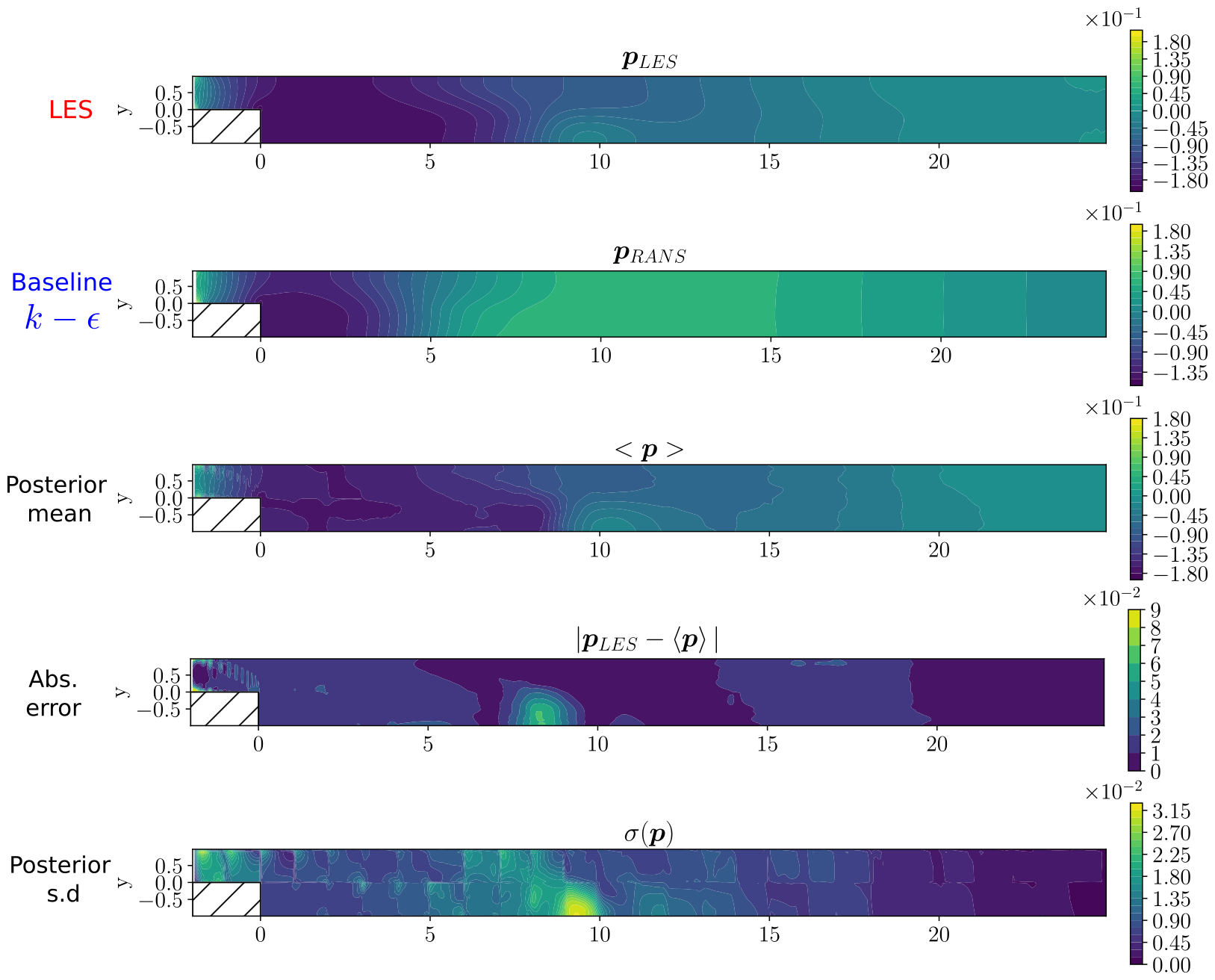}
	\caption{Pressure contours ($\bm{p}$) for the test case with Re = 500. \textbf{top} - ground truth (LES), \textbf{second} - baseline $k-\epsilon$, \textbf{third} - the posterior predictive mean ($\left<\bm{u}\right>$), \textbf{fourth} - the contour plot of the absolute error between the LES ($\bm{p}_{LES}$) and the posterior predictive mean ($\left<\bm{p}\right>$), \textbf{bottom} - the standard deviation of the posterior predictive ($\sigma(\bm{u})$).}
	\label{fig:predicted p}
\end{figure}
\begin{figure}[!t]
	\centering
	\includegraphics[width=1\textwidth]{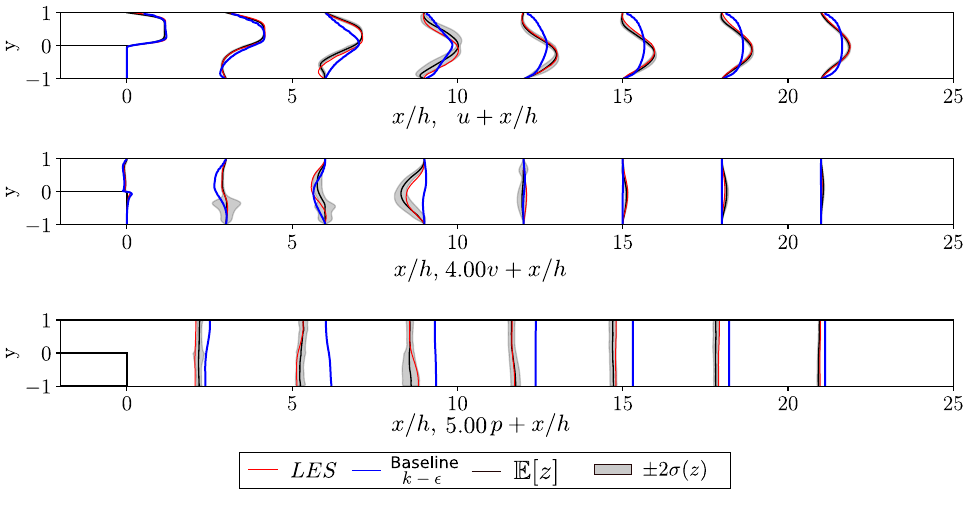}
	\caption{Section plots at different locations $x/h$ comparing the LES and \textit{Baseline} RANS mean fields with the posterior predictive mean (solid black line) and $\pm2 \times$ standard deviation (shaded area) for the test case with Re = 500.  \textbf{top} - velocity in x-direction ($\bm{u}$), \textbf{middle} - velocity in the y-direction ($\bm{v}$), \textbf{bottom} - pressure ($\bm{p}$).  $\mathbb{E}[z]$ and $\sigma(z)$ depict the predictive posterior mean and standard deviation respectively.  
	}
	\label{fig:predicted z section plots}
\end{figure}

The Monte-Carlo-based scheme (detailed in \refsec{sec:prediction}) was employed to propagate the model form uncertainty forward in order to obtain probabilistic predictive estimates for the quantities of interest i.e. mean  stream-wise velocity $u$, wall-normal velocity $v$ (\reffig{fig:predicted u}, \reffig{fig:predicted v}) and mean pressure $p$ (\reffig{fig:predicted p}).
Cross-sections of the aforementioned quantities are depicted in \reffig{fig:predicted z section plots}. The following conclusions can be drawn from the Figures:
\begin{itemize}

	\item Even though the RS field is captured with some discrepancies, the predicted mean fields agree well with the reference LES data  (Figures \ref{fig:predicted u}, \ref{fig:predicted v} and \ref{fig:predicted p}, discussed in the sequel). This points to the non-uniqueness of this inverse problem solution, also reported by other works \cite{duraisamy_perspectives_2021,brenner2022efficient}.

	\item There exist two recirculation zones in the backward-facing step flow setup. The primary one forms just after the step and the secondary appears above it for Reynolds numbers close to 400 and above, for the given expansion ratio \cite{armaly1983experimental}. As it can be seen in  \reffig{fig:predicted u}, the proposed model is  able to predict the appearance of the two recirculation zones in close agreement with the LES, whereas the baseline RANS model underestimates the size of the first recirculation zone and almost completely misses the second  one. 
	\item The last row of \reffig{fig:predicted u} depicts the predictive posterior standard deviation of the aforementioned quantities. As expected, around the shear layers (the top of the first recirculation zone and the bottom of the second recirculation zone), the uncertainty is the highest. This is even more clearly observed in the cross-sections of  \reffig{fig:predicted z section plots} which illustrate the predictive, posterior mean plus/minus two posterior standard deviations. More importantly perhaps, one observes that the predictions envelop the reference LES values in most areas. The model is extremely confident close to the inlet, as manifested by the  very tight credible interval. As one  moves further downstream and close to the first recirculation zone, the parametric closure suffers, hence the uncertainty bounds are wider to account for it. 
	The ability to quantify aleatoric, predictive uncertainty\footnote{As mentioned earlier, MAP point-estimates for the model parameters $\bt$ were used.} is one of the main advantages of  the probabilistic model proposed in contrast to the more commonly used  deterministic counterparts as well as alternatives that can only capture epistemic uncertainty.
	\item Similarly to the stream-wise velocity, predictions for the wall-normal velocity  (\reffig{fig:predicted v}) and the pressure   (\reffig{fig:predicted p})  are  in good  agreement with the reference LES values, as opposed to the baseline RANS. 
	In the first recirculation zone, the baseline $k-\epsilon$ is completely off, while the predicted values with the credible interval cover the reference LES. Also, the pressure predictions (\reffig{fig:predicted p})  identify the crucial zone where the flow reattaches to the wall (around $x/h \approx 10$), which is very difficult to predict in general. At the reattachment point, there is a transition from low-pressure in the recirculation zone to higher pressure along the wall. The posterior standard deviation at this point is also higher than in the other regions, ensuring that the reference solution is enveloped.
\end{itemize}

%


\begin{figure}[!t]
	\centering
	\includegraphics[width=\textwidth]{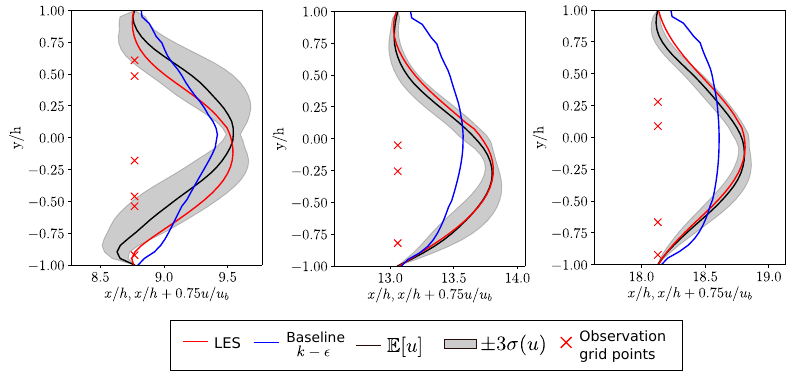}
	\caption{Section plots of the stream-wise velocity ${u}$ at three different $x/h$ locations for the test case with $Re = 500$. In addition,  the points where training observations were available are depicted with red crosses. $\mathbb{E}[u]$ and $\sigma(u)$ depict the predictive posterior mean and standard deviation respectively. 
	}
	\label{fig:u section with observables}
\end{figure}

As previously mentioned, observations of mean velocities/pressures at approximately $8\%$ of the total number of grid points in the FE mesh were used for training.  \reffig{fig:u section with observables} highlights this by comparing the stream-wise velocity at three different sections $x/h$. The left subfigure depicts the section in the first re-circulation zone. It can be seen that despite having very few  observation points near the wall, the $u$ predictions are able to capture the backward flow in the re-circulation regions. In contrast, the baseline $k-\epsilon$ completely fails to capture it. This can be attributed to the earlier reattachment of the flow in the baseline $k-\epsilon$ case (flow reattachment discussed in the sequel). 
The middle and the right  subfigures depict sections further downstream, with the middle being in the second re-circulation zone and the right in the flow recovery zone. It is observed that with a relatively small number of observation points, the trained model's  prediction is in agreement with the LES. Furthermore, the latter is enveloped by the credible interval constructed by considering  $\pm 3\times$ (the posterior standard deviation). This credible interval is much tighter as compared to the one in the left subfigure.

%
\update{
The primary motivation behind RANS models and of this work is to reduce the computational cost of flow simulations, especially in cases  where  LES or DNS are prohibitively expensive. In Table \ref{table:computational_cost} we report the computational time of a single, model run as compared to the baseline RANS model(for $Re=500$).
We observe that even though the proposed differentiable solver is of comparable cost to the baseline RANS, it can provide far more accurate, probabilistic predictions of the LES model's outputs which is  roughly $6000$ times more computationally demanding.
}


\begin{table}[!htbp]
	\centering
	\begin{tabular}{p{2.5cm} c c}
		\toprule 
		  & LES & proposed RANS solver\\
		\hline\\ 
	   Relative cost & 7679 & 1.3  \\
		\bottomrule
	\end{tabular}
	\caption{\update{Relative computational cost in terms of computational time (on an Intel Core i9-12900K CPU) of a single model run for LES and of the proposed, differentiable RANS solver.  These are compared with the cost of the baseline RANS model for Reynolds number $500$. 
	}}
	\label{table:computational_cost}
\end{table}

Accurately capturing the recirculation zones is crucial to getting a reliable estimate of the reattachment length, which is a key parameter in the study of separated flows, such as the case  here. The reattachment length is defined as the distance from the step where the flow separates to the point at which it reattaches 
to the surface downstream of the step. Reattachment occurs where the velocity gradient off the wall is zero, or in other words, where the wall shear stress is zero. The predicted  reattachment length ($x_{reattach}$) by  the proposed method is compared with a)the  LES data (\refsec{sec:training_data}) b) the  baseline RANS (\refsec{sec:training_data}) c) the two-dimensional,  LES simulation performed by \cite{biswas2004backward} for the expansion ratio of $1.9423$,  and d) the results in  \cite{geneva2019quantifying}, who used the same Tensor Basis Neural Networks (TBNN) \cite{ling2016reynolds} employed in the parametric closure term in our work as well. The results are summarized in Table \ref{table:Reattachmenet length} where it is evident that while previous works deviated significantly from the reference LES value, our probabilistic prediction is able to  envelop it. 
The reattachment length is heavily dependent on correctly identifying the two recirculation zones and the baseline RANS model fails to predict the secondary recirculation zone (\reffig{fig:predicted u}). This might have resulted in such a low reattachment length. The estimate of the reattachment length is also low in \cite{geneva2019quantifying}, which could  be attributed to the lack of the stochastic, model correction term. 


\begin{table}[!htbp]
	\centering
	\begin{tabular}{l l} 
		\hline \\[-1.5ex]
		Model & $x_{reattach}$[x/h]\\[0.5ex]
		\hline \\[-1ex]
			LES (reference) & 9.10\\
		Biswas et al. \cite{biswas2004backward} ($H/h = 1.9423$) & 8.9\\
		Baseline RANS ($k-\epsilon$ model) & 5.61\\
		Geneva et al.\cite{geneva2019quantifying} & 5.52\\
			proposed  model & 10.06 $\pm$ 1.21\\
			[0.5ex]
		\hline
	\end{tabular}
	\caption{Predictions of reattachment length ($x/h$) of the primary recirculation region behind the backward-facing step (expansion ratio $H/h=2$) for the test case with $Re = 500$. The $\pm$ corresponds to $3 \times$ the posterior standard deviation. 
	}
	\label{table:Reattachmenet length}
\end{table}


\section{Conclusions}
\label{sec:conclusion}
We have presented a data-driven  model for RANS simulations that quantifies and propagates in its predictions an often neglected source of uncertainty, namely the aleatoric, model uncertainty in the closure equations. We have combined this with a parametric closure model which employs  a set of tensor basis functions that depend on the invariants of the rate of  strain and rotation tensors. A fully Bayesian formulation is advocated which makes use of  a sparsity-inducing prior   in order to identify the regions in the problem domain  where the parametric closure is insufficient and in order to  quantify the stochastic correction to the Reynolds stress tensor.
We have demonstrated how the model can be trained using sparse, indirect data, namely mean velocities/pressures in contrast to the majority of pertinent efforts that require direct, RS data. While the training data in our illustrations arose from a higher-fidelity model, one can readily envision using experimental observations as well.

In order to enable inference and learning tasks, we  developed a differentiable RANS solver capable of providing parametric sensitivities.  Such a differentiable solver was non-trivial owing to the complexity of the physical simulator and its stability issues. The lack of such numerical tools has proven to be a significant barrier for intrusive, physics-based, data-driven  models in  turbulence \cite{cranmer2020frontier}. This differentiable solver was utilized in the context of a  Stochastic Variational Inference (SVI) scheme that employs Monte Carlo estimates of the ELBO derivatives in conjunction with the reparametrization trick and stochastic gradient ascent. We demonstrated how probabilistic predictive estimates can be computed  for all output quantities of the trained RANS model and illustrated their accuracy on a separated flow in the backward facing step benchmark problem. In most cases, very good agreement with the reference values was achieved and in all cases these were enveloped by the credible intervals computed.

The proposed modeling framework offers several possibilities for extensions, some of which we discuss below:

\begin{itemize}
	\item The indirect data i.e. velocities/pressures as in the \refeq{eq:likelihood}, could be complemented with direct, RS data at certain locations of the problem domain. This could be beneficial in improving the model's predictive accuracy and generalization  capabilities.
	\item The parametric closure model could benefit from non-local dependencies  which could be enabled  by  convolutional or  vector-cloud neural networks (VCNN)  \cite{han2022vcnn}  with appropriate embedding of invariance properties. 
\item The dimensionality reduction of the stochastic discrepancy terms (\refeq{eq:epsdim}) was based on a pre-selected and uniform division of the problem domain into subdomains. The accuracy of the model would certainly benefit from a learnable and adaptive scheme that would be able to focus on the areas where model deficiencies are most pronounced and stochastic corrections are most needed.

\end{itemize}

\section*{Acknowledgements}
\noindent We extend our sincere thanks to Jigar Parekh for his valuable guidance in the generation of the LES data and to Maximilian Rixner for his assistance in implementing the forward solver. We are also thankful to Nicholas Geneva and Niklas Fehn for their valuable discussions that enriched this work. Our sincere gratitude is extended to the anonymous reviewers, whose insightful feedback greatly enhanced the quality of our manuscript.

\clearpage

\appendix

\section{Differentiable RANS solver}
\label{app:RANS solver details}

In the present study, the RANS equations (\refeq{eq:RANS_equation}) are numerically solved using the finite element discretization, implemented in the open source package FEniCS \cite{alnaes2015fenics}. The discrete equations are obtained by representing the solution and test functions in appropriate finite dimensional function spaces. In particular, we employed the standard Taylor-Hood pair of basis functions with  polynomial degree one for the pressure interpolants and two for the  velocities. This choice is made to avoid stability issues potentially arising from the interaction between the momentum and continuity equations. 

The turbulence scaling terms, $k$ and $\epsilon$ in \refeq{eq:TBNN_model}, are obtained by solving the respective standard transport equations \cite{pope2000turbulent,alfonsi_reynolds-averaged_2009}. Symmetry is enforced in the RS tensor, i.e. $\tau_{xy}$ and $\tau_{yx}$ are identical without any redundancy in the representation. The discretized system is solved with damped Newton's method. For robustness and global convergence, pseudo-time stepping is used with the backward Euler discretization \cite{deuflhard2005newton}.
As the Reynolds number is increased, the convection term dominates, leading to stability \cite{donea2003finite}.
This elicits a need to add stabilization terms to the weak form, such as the least-square stabilization, according to which the weighted square of the strong form  is added to the weak form residual. However, these extra terms have to be chosen carefully in order not to compromise the correctness of the approximate solution. Classically, researchers added artificial diffusion terms or a numerical diffusion by using upwind scheme for the convection term instead of central diffusion. The extra infused term corrupted the solution quality. To avoid this, in practice, it is common to use schemes like Streamline-Upwind Petrov-Galerkin method (SUPG) and Galerkin Least Squares (GLS). In the present study, we have utilized a self-adjoint numerical stabilisation scheme which is an extension of Galerkin Least Squares (GLS) Stabilisation called Galerkin gradient least square method \cite{franca1989galerkin}. This amounts to adding a stabilization term to the residual weak form. For additional details, interested readers are referred to \cite{franca1989galerkin,donea2003finite}.

\section{ Adjoint Formulation and Estimation of the Gradient of the ELBO}
\label{app:grad ELBO}
%
As discussed in Section \ref{sec:learning}, the SVI 
framework advocated, in combination with the reparametrization  trick, requires derivatives of the ELBO with respect to the variables which we summarily denoted by $\bm \eta = \{\bm \theta,\bm \Lambda, \bm{E}_{\tau}^{(1:M)}\}$, i.e. (as in \refeq{eq:ELBO_grad_est_cont}):
\be
	\nabla_{\bm \xi} \calF(\bm \xi)  = \mathbb{E}_{q(\bm \phi)} \left[ 
	\nabla_{\bm{\xi}}g_{\xi}(\bm \phi)\nabla_{\bm \eta}
	\left(\log  p ( \mathcal{D},\bm{\eta})- 
	\log q_{\xi} (\bm{\eta})\right)   \right],
	\ee
where from \refeq{eq:post}:
\be
\begin{array}{ll}
\log  p ( \mathcal{D},\bm{\eta} )  & = \log \left( p(\mathcal{D}~ |~ \bt, \bm{E}_{\tau}^{(1:M)}) ~p( \bm{E}_{\tau}^{(1:M)} | \bm{\Lambda})~p(\bt)  ~p(\bm{\Lambda})  \right)\\
	& = \left(   \sum_{m=1}^M \log p(\zobs^{(m)} \mid \bt, \bm{E}_{\tau}^{(m)}) + \log p(\bm{E}_{\tau}^{(m)}| \bm{\Lambda}) \right)\\
 &~+\log p(\bt)  +\log ~p(\bm{\Lambda}).
\end{array}
\label{eq:logjoint}
\ee
The form of the (log-)priors $p(\bm{E}_{\tau}^{(m)}| \bm{\Lambda})$ (\refeq{eq:gprior}), $ p(\bt)$ (\refeq{eq:prior_theta}), $p(\bm{\Lambda})$ (\refeq{eq:lprior}) as well as of the approximate posterior $q_{\xi} (\bm{\eta})$ (Equations (\ref{eq:qmean}) and (\ref{eq:posttheta}), (\ref{eq:postlambda}), (\ref{eq:poste})) suggest that most of these derivatives can be analytically computed with the exception of the  ones involving the log-likelihoods, i.e.:
\be
\ell^{(m)} ( \bt, \bm{E}_{\tau}^{(m)})=\log p(\zobs^{(m)} \mid \bt, \bm{E}_{\tau}^{(m)}).
\label{eq:loglikem}
\ee
This is because each of these terms  depends implicitly on $\bt, \bm{E}_{\tau}^{(m)}$ through the output of the RANS solver $\z( \bt, \epsTau^{(m)}=\bm{W}\bm{E}_{\tau}^{(m)})$ with the closure model for the discretized RS tensor field suggested by \refeq{eq:taudecomp} i.e.  $\bm{\tau}=\bm{\tau}_{\bt}(\bm{u})+ \bm{W}\bm{E}_{\tau}^{(m)}$. In view of the governing equations (\refeq{eq:residual_tau_theta}), we explain below how adjoint equations can be formulated that enable efficient computation of the aforementioned derivatives of the log-likelihoods.

In particular, and if we drop the superscript $m$ for each term in the log-likelihood in order to simplify the notation, we formulate a Lagrangian with the help of a vector  $\bm \lambda$ of Lagrangian multipliers, i.e.:
\begin{align}
	\calL=\ell + \bm \lambda^T(\calG(\z)-\bm{B} \bm{\tau} ),
\end{align}
where $\calG$, $\bm{B}$, $\taub$ and $\z$ are as defined in \refsec{sec:methods}.Differentiating with respect to $\taub$ yields:
\begin{align}
	\frac{d\calL}{d\bm{\tau}}&= \frac{\partial \ell}{\partial \z } \frac{d \z}{d\bm{\tau}} + \frac{d \bm \lambda^T}{d\bm{\tau}} (\calG(\z)-\bm{B} \taub) + \bm \lambda^T \left(\frac{\partial \calG}{\partial \z} \frac{d \z}{d\bm{\tau}} - \bm{B} \right)\nonumber \\
	&= \left(\frac{\partial \ell}{\partial \z} + \bm \lambda^T\frac{\partial \calG}{\partial \z} \right)\frac{d \z}{d \bm{\bm{\tau}}} - \bm \lambda^T \bm{B} .
	\label{eq:adjoint_expanded}
\end{align}
We select $\bm \lambda^T$ so that the first term in parentheses vanishes, i.e. :

\begin{align}
	\frac{\partial \ell}{\partial \z} + \bm \lambda^T\frac{\partial \calG}{\partial \z} = 0 \quad \textrm{or,} \quad \left(\frac{\partial \calG}{\partial \z} \right)^T \bm \lambda = - \left(\frac{\partial \ell}{\partial \z} \right)^T.
\end{align}
%
The linear system of equations was  solved using  a direct LU solver. The vector $\bm \lambda$ found was substituted  in \refeq{eq:adjoint_expanded} in order to obtain the desired gradient which is  given by:
\begin{align} \label{eq:Adjoint_gradient}
	\frac{d\calL}{d\bm{\tau}} =\frac{d\ell }{d\bm{\tau}} = - \bm \lambda^T \bm{B}. 
\end{align}

Subsequently, and by application of the chain rule we can obtain derivatives with respect to $\bt$ as:
\begin{align}\label{eq:grad_wrt_theta}
	\frac{d \ell}{d \bm \theta} = \underbrace{\frac{\pa  \ell}{\pa \taub}}_{\substack{\text{Adjoint}\\ \text{model}}} \underbrace{\frac{\pa \taub}{\pa \bm \theta}}_{\substack{\text{NN}\\ \text{auto-diff}}},
\end{align}
where  $\pa \taub/\pa \bm \theta$ was efficiently  computed by back-propagation, which is a reverse accumulation automatic differentiation algorithm for deep neural networks that applies the chain rule on a per-layer basis.
We note that since the parameters $\bt$ are common for each likelihood $\ell^{(m)}$ the aforementioned terms would need to be added as per \refeq{eq:logjoint}.

Similarly by chain rule,  the gradient with respect to the vector $\ETau$ is given by:
\be
	\frac{d \ell}{d \ETau  }= \bm{W}^T \frac{d \ell}{d \taub}.
	\label{eq:grad_wrt_ETau}
\ee
We note finally that the expectations involved in the ELBO and its gradient (\refeq{eq:ELBO_grad_est_cont})   are  approximated by Monte Carlo i.e.:
\begin{align}\label{eq:elbomc}
\mathcal{F}(\bxi)\approx \frac{1}{K} \left( \sum_{k=1}^K \left( \sum_{m=1}^M \ell^{(m)} (\bt, \bm{E}_{\tau}^{(m,k)}) + \log p(\bm{E}_{\tau}^{(m,k)} |  \bm{\Lambda}) \right)  
~+\log p(\bt) +\log ~p(\bm{\Lambda}) - 	\log q_{\bm \xi} (\bt,\bm{\Lambda},\bm{E}_{\tau}^{(m,k)})\right),
\end{align}
where:
\be 
\bm{\phi}^{(m,k)}\sim \mathcal{N}(\bm{0,I}), ~~\bm{E}_{\tau}^{(m,k)}=g_{\bm{\xi}}(\bm{\phi}^{(m,k)}),
\ee
and:
\be
\nabla_{\bm \xi} \calF(\bm \xi) \approx
	\frac{1}{K} \sum_{k=1}^{K}
	   \nabla_{\bm{\xi}}g_{\xi}(\bm \phi^{(k)})\nabla_{\bm \eta}
	\left(\log  p ( \mathcal{D},\bm{\eta}^{(k)})- 
	\log q_{\xi} (\bm{\eta}^{(k)})\right),\\
	\label{eq:elbodermc}
\ee
where $\bm \eta^{(k)} = g_{\xi}(\bm{\phi}^{(k)})$.
\bibliographystyle{elsarticle-num}
\bibliography{bibliography}



\end{document}